\documentclass[journal]{IEEEtran}
\ifCLASSINFOpdf
\else
\fi

\usepackage{tipa} 
\usepackage[scaled=0.95]{inconsolata}  

\ifCLASSOPTIONcompsoc
  \usepackage[caption=false,font=normalsize,labelfont=sf,textfont=sf]{subfig}
\else
  \usepackage[caption=false,font=footnotesize]{subfig}
\fi

\usepackage[ruled,vlined,linesnumbered]{algorithm2e}
\usepackage{amsmath,graphicx}
\usepackage{bm}
\usepackage{color}
\usepackage{amssymb}
\usepackage[hyphens]{url}
\usepackage{hyperref}
\hypersetup{
    colorlinks=true,
    linkcolor=blue,
    urlcolor=magenta,
}
\usepackage{mwe}
\usepackage{booktabs}
\usepackage[pagewise,switch]{lineno}
\usepackage[space]{cite}
\usepackage{float}

\usepackage{footnote}
\makesavenoteenv{tabular}

\usepackage{pifont}
\newcommand{\cmark}{\ding{51}}%
\newcommand{\xmark}{\ding{55}}%

\usepackage{array}
\newcolumntype{P}[1]{>{\centering\arraybackslash}p{#1}}

\usepackage{lipsum}

\usepackage{lscape}
\usepackage{multirow}
\usepackage{makecell}
\usepackage{pdflscape}
\begin{document}
%
\title{Vibravox: A Dataset of French Speech Captured with Body-conduction Audio Sensors}
%
%
%

\author{Julien~Hauret,
        Malo~Olivier,
        Thomas~Joubaud,
        Christophe~Langrenne,
        Sarah~Poirée,
        Véronique Zimpfer,
        and Éric~Bavu
\thanks{Julien Hauret, Malo~Olivier, Christophe~Langrenne, Sarah~Poirée, and Éric~Bavu are with the Laboratoire de Mécanique des Structures et des Systèmes Couplés, Conservatoire national des arts et métiers, HESAM Université, 75003 Paris, France. e-mail for these authors: name.surname@lecnam.net.\\
    Thomas~Joubaud and Véronique~Zimpfer are with the Department of Acoustics and Soldier Protection, French-German Research Institute of Saint-Louis (ISL). e-mail for these authors: name.surname@isl.eu}
\vspace{-4mm}}

%
%

\markboth{Submitted to Elsevier Speech Communication on August 20, 2024 - Revision submitted on March 29, 2025}%
{}
%



\maketitle

%
%
%


\begin{abstract}
Vibravox is a dataset compliant with the General Data Protection Regulation (GDPR) containing audio recordings using five different body-conduction audio sensors: two in-ear microphones, two bone conduction vibration pickups, and a laryngophone. The dataset also includes audio data from an airborne microphone used as a reference. The Vibravox corpus contains 45 hours per sensor of speech samples and physiological sounds recorded by 188 participants under different acoustic conditions imposed by a high order ambisonics 3D spatializer. Annotations about the recording conditions and linguistic transcriptions are also included in the corpus. We conducted a series of experiments on various speech-related tasks, including speech recognition, speech enhancement, and speaker verification. These experiments were carried out using state-of-the-art models to evaluate and compare their performances on signals captured by the different audio sensors offered by the Vibravox dataset, with the aim of gaining a better grasp of their individual characteristics.
\end{abstract}

\begin{IEEEkeywords}
Body-Conduction audio sensors, Robust Communication, Speech enhancement, Speech recognition, Speaker verification
\end{IEEEkeywords}
\vspace{-1mm}

%
\IEEEpeerreviewmaketitle

%
%
%
%
\vspace{-5mm}
\section{Introduction}
\label{sec:intro}

\IEEEPARstart{U}{nlike} traditional microphones, which rely on airborne sound waves, body-conduction audio sensors -- often referred to as body conduction microphones (BCMs) for simplicity -- allow voice signals to be picked up directly from the body, offering potential advantages in noisy environments by greatly reducing the influence of ambient noise on recordings. Although BCMs have been available for decades \cite{graciarena2003combining,zheng2003air,bos2005speech,acker2005speech,shahina2007mapping}, the limited bandwidth of speech captured by such sensors has so far restricted their widespread use. However, thanks to two tracks of improvements, this technology could be made available to a wider audience for speech capture and communication in noisy environments.

Research into physics and electronics is improving with some skin-attachable sensors such as \cite{lee2019ultrathin,lee2022electret,che2024speaking}. Similarly to earlier bone and throat microphones, these new wearable sensors detect skin vibration, which is highly and linearly correlated with the acoustic pressure produced by the wearer's voice \cite{bjorklund2016relationship}. Not only do they improve on the state of the art by having superior sensitivity over the vocal frequency range -- thereby improving the signal-to-noise ratio -- but they also have superior skin compliance, which facilitates adhesion to curved skin surfaces. These new kinds of sensors, just like the previous ones, are, however, unable to capture the full bandwidth of the speech signal due to the inherent low-pass filtering of tissues. The manufacturing process also needs to be stabilized, which currently prevents them from being commercially available. In addition, some earbuds, such as the Xiaomi Buds 4 Pro, already include some voice pickup unit thanks to a retro-action microphone integrated into the loudspeaker at the entrance to the ear canal. Finally, the innovative Aria Gen 2 Meta's research platform now integrates a contact microphone. This exemplifies the significant role that BCMs will likely assume in the future.

Deep learning methods have shown excellent performance in a variety of speech and audio tasks. In this work, we show that the Vibravox dataset can be used to advance research in three key tasks that would enable overcoming the current limitations of body conduction sensors: bandwidth extension, speech recognition, and speaker verification. Since body conduction sensors exhibit reduced bandwidth transduction efficiency, works such as \cite{zhang2021wsrglow, han2022nu,serra2022universal, mandel2023aero,nistico2023audio,shuai2023mdctgan, hauret2023eben, hauret2023configurable, karthikeyan2023speech, andreev2023hifi++,li2023restoration,he2023towards, li2023two, edraki2024speaker,ohlenbusch2024multi,sui2024tramba,yu2024bae,scheibler2024universal,liu2024audiosr} demonstrate the ability of deep learning approaches to regenerate mid and high frequencies from low-frequency audio content. They have adapted recent deep learning trends to the specific problem of bandwidth extension, also known as audio super-resolution. For robust speech recognition, models such as Whisper \cite{radford2022whisper} or Canary1B \cite{canary1b} have pushed the limits of usable signals. Finally, for speaker verification, approaches such as TitaNet \cite{koluguri2022titanet}, WavLM \cite{chen2022wavlm}, Pyannote \cite{bredin2023pyannote}, and ECAPA2 \cite{thienpondt2023ecapa2} now allow a wide range of signals to be used thanks to their increased robustness.

The availability of large datasets is critical to advancing research and development of BCMs. These datasets, allowing the training and evaluation of deep learning models, have been a key missing ingredient in achieving high-quality, intelligible speech with such audio sensors. The assumption that airborne and body-transmitted speech share identical excitation sources and possess a simple transfer function between them is inadequate; producing realistic synthetic data is therefore still a challenging task. In a previous study, we performed out-of-distribution tests performed on real BCM signals with an EBEN model trained on simulated data \cite{hauret2023configurable}, showing the importance of using a sufficiently large set of real data for training. Despite their importance, existing datasets still exhibit several notable limitations. Data collection being labor-intensive, the existing BCM datasets often lack the necessary scale and diversity to comprehensively cover the full range of acoustic scenarios encountered in real-world applications. Another limitation is the absence of publicly available BCM datasets in French, restricting research primarily to the languages covered by existing resources. Problems also persist with signal quality, including noise and artifacts. On top of these issues, the variety of sensors used in existing datasets remains limited, making it difficult to generalize results across different recording environments and equipment configurations. The Vibravox dataset is being made available to fill these gaps and stimulate research in the field of speech capture using non-conventional audio sensors.

\begin{table*}[ht!]
\caption{Open source BCM Dataset review}
\vspace{-2mm}
\centering
\begin{tabular}{@{}cP{1.35cm}P{1.25cm}P{1.3cm}P{1.3cm}P{1.3cm}cP{1.25cm}P{1.25cm}c@{}}
\toprule
\textbf{Dataset} & \textbf{Number of speakers}& \textbf{Number of BCM} & \textbf{Clean speech}  & \textbf{Noisy speech} & \textbf{Speechless recordings}  & \textbf{Language} & \textbf{Text transcript} & \textbf{Phonetic transcript} & \textbf{Download page} \\
\midrule
      ABCS  \cite{wang2022abcs} &  100 & 1 & $2 \times 42$h & 0h  & 0h & Chinese & \textcolor{green}{\cmark} & \textcolor{red}{\xmark} &  \href{https://github.com/elevoctech/ESMB-corpus}{GitHub}\\
      ESMB \cite{esmb2021repo}  & 287 & 1 & $2 \times 128$h & 0h& 0h & Chinese & \textcolor{red}{\xmark} & \textcolor{red}{\xmark} & \href{https://github.com/wangmou21/abcs}{GitHub}\\
      EmoBone \cite{hosain2024emobone} & 28 & 1 & $2 \times 19$h & 0h & 0h & English & \textcolor{green}{\cmark} & \textcolor{red}{\xmark} & Avail. upon request \\
      TAPS \cite{kim2025taps} & 60 & 1 & $2 \times 13$h$18$ & 0h & 0h & Korean & \textcolor{green}{\cmark} & \textcolor{red}{\xmark} & \href{https://huggingface.co/datasets/yskim3271/Throat_and_Acoustic_Pairing_Speech_Dataset}{HuggingFace} \\
\midrule
      Vibravox (ours)  \cite{cnamlmssc2024vibravoxdataset}  &  188  & 5 & $6 \times 33$h$18$ & $6 \times 1$h$58$ & $6 \times 10$h$21$ & French & \textcolor{green}{\cmark} & \textcolor{green}{\cmark} & \href{https://huggingface.co/datasets/Cnam-LMSSC/vibravox}{HuggingFace}\\
\bottomrule
\end{tabular}
\label{tab:datasets_review}
\end{table*}

In this paper, we contribute along two major axes: the construction and utilization of the dataset. First, we introduce the Vibravox dataset, developed through a rigorous data collection process. We designed and implemented a recording protocol, assembled the necessary hardware, and built dedicated software to ensure a controlled and high-quality data acquisition setup. A total of 200 participants contributed to the dataset, with 188 retained after post-processing to ensure high-quality data. We applied post-filtering techniques to enhance data quality and facilitate interpretability, enabling meaningful comparisons between different microphones and insightful speech signal analyses. Second, we establish baselines using the Vibravox dataset for three key speech processing tasks: bandwidth extension, speech-to-phonemes recognition, and speaker verification. For each task, we conduct experiments in both quiet and noisy environments to comprehensively evaluate the strengths and limitations of body-conduction microphones under realistic conditions. For speech-to-phonemes recognition and speaker verification, we compare the performance of raw and enhanced body-conduction sensors in quiet conditions to evaluate the impact of speech enhancement on downstream applications. These contributions provide new insights into the potential and challenges of BCMs for speech capture and processing, paving the way for future advancements in the field.

\section{Related work}
\label{sec:related_work}

The need for comprehensive and publicly available datasets of own-speech recordings using BCMs, such as those listed in Table \ref{tab:datasets_review}, is critical to the progress of research and development in the field of body-conducted speech capture. These datasets are essential for training and testing deep learning models.

\subsection{Training datasets}

Before delving into publicly available datasets, it is worth acknowledging that they have been made available for a relatively short period of time, are in some cases of insufficient size, and are only available in a restricted set of languages. Consequently, several studies have employed low-pass filtering of high-quality audio to artificially generate body-conducted speech data. A hybrid approach involves the collection of individual transfer functions between a reference microphone and a body-conducted microphone, which is the focus of the Hearpiece database, as referenced in \cite{denk2021hearpiece}. In their article, Ohlenbusch et al. \cite{ohlenbusch2023modeling} proposed several speech-dependent models of one's own voice transfer function to simulate the degradation induced by body-conducted audio recordings. Their results indicate that the transfer function is, of course, speaker-dependent, but also phoneme-dependent due to the different position of the jaw. Their study brings a significant improvement in the ability to simulate data. Following on from this work, the authors also published \cite{ohlenbusch2024speech} to propose speech-dependent data augmentation to compensate for the lack of a large dataset of own speech signals. They were able to reduce the performance gap when testing the model on real signals. However, by analogy with FineWeb-Edu \cite{lozhkov2024fineweb-edu}, the performance improvement can also come from having a larger and highly curated dataset on which to perform classical supervised training.

A comprehensive literature review of existing, publicly available datasets of BCM-captured speech is essential to gain a full understanding of the subject. This encompasses all forms of capture, including bone, in-ear, and throat sensors. It should be noted that there exist several small private datasets \cite{erzin2009improving,turan2015source,shan2018novel,davis2019ear, tagliasacchi2020seanet,zheng2022dual, schilk2023ear}, albeit not open-sourced. The few remaining public datasets are listed in Table \ref{tab:datasets_review}, including Vibravox. In terms of size, the ESMB dataset \cite{esmb2021repo} is the largest, with 128 hours of recorded speech, followed by Vibravox with its 45h37 of recordings per sensor, ABCS \cite{wang2022abcs} (42 hours), and EmoBone \cite{hosain2024emobone} (19 hours). The final dataset to be considered is TAPS (13h18) \cite{kim2025taps}. This dataset includes post-recording processing to realign signals at the sample level, thereby easing the direct utilization of L1 loss in the waveform domain for the training of deep learning models to enhance body-conducted speech. Although Vibravox is not the largest dataset, it overcomes several limitations found in existing datasets, such as the limited diversity of audio sensors and the lack of noise recordings. Vibravox addresses these issues by using five different BCMs and including recordings of speech in both noisy and quiet environments. Previous studies indicate that the transmission of external noise through the device is not only influenced by the device itself but also by individual variance \cite{denk2021hearpiece} and angle of arrival \cite{liebich2018direction}. For this reason, the Vibravox dataset was recorded in a 3D sound spatializer to sample uniformly the sphere for noise emission. Its unique combination of simultaneous multi-microphone recordings and French-language data therefore provides complementary resources to existing datasets, offering new perspectives for non-conventional speech capture and comparative analyses of different sensor technologies.

\subsection{Speech processing tasks}

The body-conducted speech research community has made significant progress in recent years. The accessibility of public datasets such as Vibravox has become increasingly important for the development and improvement of deep learning models in this field. The continued availability of these resources is likely to facilitate further advances and innovations in body-conducted speech technology for tasks such as speech enhancement, speech transcription, and speaker verification.\\

\subsubsection{Speech Enhancement}
BCM datasets, paired with airborne speech, are invaluable in the field of speech enhancement. This task refers to the process of improving the quality and intelligibility of speech, often in the presence of background noise. However, speech enhancement on BCM is mainly about bandwidth extension, as the sensors are robust to noise but cannot record high frequencies due to physical constraints. Several approaches have been published recently in this context. Among them, \cite{zheng2022dual,li2023restoration,sui2024tramba} model the long-term dependencies of speech by adapting the non-quadratic complexity variations of the attention mechanism \cite{vaswani2017attention,liu2021swin,gu2023mamba}. Some studies show superior reconstruction metrics compared to EBEN, but as shown in Table 1 of \cite{hauret2023eben}, one of the approaches reporting the best metrics is Kuleshov's U-net \cite{kuleshov2017audio}, which is also the one that performs worst in the MUSHRA (MUltiple Stimuli with Hidden Reference and Anchor) analysis. This evidence serves to illustrate the importance of conducting listening tests when comparing reconstructive and generative approaches. Furthermore, EBEN's computational and memory requirements are low, due not only to the limited number of parameters in the generator but also to the highly strided embeddings facilitated by the Pseudo-Quadrature Mirror Filters (PQMF) representation \cite{hauret2023configurable}. In \cite{edraki2024speaker}, the authors proposed an innovative approach to deal with limited datasets. They developed a method to disentangle a speaker embedding vector from the hidden representations of the network during the enhancement process, which allowed a fair generalization performance with a limited training set of 5 hours of 29 speakers and additional synthetic data. Conversely, the approach proposed in \cite{sui2024tramba} involves pre-training with simulated data and fine-tuning on a device with real data, thereby creating a speaker-dependent speech enhancement model.\\

\subsubsection{Speech-to-text}
Transcribed speech datasets are essential for training models to accurately recognize and transcribe spoken words. Eventually, performances can be boosted with a self-supervised pretraining \cite{oord2018representation} on unlabeled speech data to construct a semantic representation of speech prior to using text labels. To perform robust transcription, the inclusion of speech in both noisy and quiet environments in the Vibravox dataset is particularly beneficial, as it allows for the development of models that can perform well in a variety of acoustic conditions. To provide context, \cite{ishimitsu2010body,wang2022abcs,cao2023can} have previously addressed this task using this particular data modality. They have shown that body-conducted speech can improve the performance of a basic speech recognition system in adverse environments solely based on airborne speech.\\

\subsubsection{Speaker verification}
Finally, speaker verification is a biometric authentication method that uses a person's voice to confirm their identity. As with all the other tasks, if the microphone is less affected by reverberation and external noise, the problem of domain mismatch is reduced. This, in turn, enables higher recognition accuracy in challenging environments. Another benefit that is not immediately apparent is the amount of information available. Because everyone's skull and timbre are unique, the recorded signal can be more descriptive of a person's identity. This wealth of biometric information opens up several potential applications, including those in the military domain, as detailed in \cite{everett1985automatic}. One such application is access control, where it can strengthen security by adding an extra layer compared to conventional methods like keycards or codes. Another compelling application lies in intelligence gathering. By pinpointing known voices in intercepted transmissions, speaker identification can reveal the involvement of high-value targets. Beyond this paper's focus, this rich biometric data (heart rate, breathing) could be a game-changer in battlefield forensics, aiding in revealing emotional states and tracking life signals during critical events.

\section{Building Vibravox}

Creating the Vibravox BCM dataset involved a number of different tasks. These included designing pre-amplification and conditioning circuits for each transducer, coding the front-end and back-end software for the recording user interface, adapting the housings and fixings for some audio sensors to the human body through 3D modeling and printing, and selecting transducer technologies and recording parameters. For the sake of brevity, this section will only discuss the essential parts needed to gain a clear understanding of the collected data. All other details can be found at the following URL: \url{https://vibravox.cnam.fr/}, including pictures of the actual recording environment as well as some fully equipped participants during recording sessions.

\begin{figure}[htb]
  \centering
  \centerline{\includegraphics[width=0.98 \linewidth]{fig_idx/fig_01.pdf}}
\caption{Fully equipped participant with the six audio sensors}
\label{fig:headset}
\end{figure}

\begin{table*}[h]
     \caption{Sensors specifications}
\vspace{-2mm}
\centering
\resizebox{2.1\columnwidth}{!}{%
\begin{tabular}{@{}cccccc@{}}
\toprule
\textbf{Name} & \textbf{Location} & \textbf{Technology} & \textbf{Reference} & \textbf{Signal-to-self-noise} & \textbf{Signal-to-external-noise}  \\
\midrule
  	  Headset microphone &  Close to mouth & Cardioid electrodynamic microphone & Shure WH20XLR & 36.0 dB & 0.1 dB \\
  	  Forehead accelerometer &  Forehead & One-axis piezoceramic accelerometer  &  Knowles BU23173-000  & 23.7 dB  & -3.5 dB \\
  	  In-ear soft foam microphone &  Left ear & Omni. electret condenser microphone &  Knowles FG-23329-P07 & 22.8 dB  & 4.0 dB\\
      In-ear rigid earpiece microphone &  Right ear & Omni. MEMS microphone  & Knowles SPH1642HT5H & 29.0 dB  & 3.9 dB\\
      Throat microphone &  Larynx & Piezoelectric  & iXradio XVTM822D-D35 & 41.1dB  & 25.6 dB\\
      Temple vibration pickup &  Temple & Figure of-eight condenser transducer  &  AKG - C411 & 9.5 dB  & 8.9 dB \\
\bottomrule
\end{tabular}}
\label{tab:specifications}
\end{table*}

\subsection{Hardware}
\label{subsec:hardware}

A comprehensive list of all BCMs employed in the Vibravox dataset is presented in Tab.~\ref{tab:specifications}. Their positioning on the participants' heads is shown in Figure \ref{fig:headset}. Although not depicted in Figure \ref{fig:headset}, a custom 3D-printed backpack and headset were also designed with the objective of facilitating the optimal placement of sensors, carrying the Zoom F8n field recorder, and routing audio cables to that device\footnote{Several pictures and a video of participants wearing the sensors and the backpack during experiments are available at \url{https://vibravox.cnam.fr/documentation/hardware/sensors}}. The Zoom F8n records all six sensor signals simultaneously, operates at a sampling rate of 48 kHz and a resolution of 32 bits, thereby guaranteeing precise signal capture with high fidelity. To enhance audio data quality, a 20 Hz high-pass filter is uniformly applied to each acquisition channel, effectively attenuating low-frequency noise and artifacts.\\

In order to provide a balanced comparison between body conduction and airborne audio sensors in terms of immunity to ambient noise, we selected an airborne microphone mounted on a headset that presents a cardioid directivity pattern towards the speaker's mouth, hence producing a dry reference. The in-ear soft foam-embedded microphone is a prototype designed by the ISL and used in \cite{blonde2023numerical}. The in-ear rigid earpiece-embedded microphone was initially presented in \cite{denk2019one} as a versatile research earpiece with multiple drivers and microphones. In this study, we use the occluded version of this device and only record the in-ear microphone signal. The laryngophone has been purchased from iXradio. This device consists of dual piezoelectric sensors on an adjustable neck rim, which captures skin vibrations from vocal cords and targets voice amplifiers, recorders, and Voice over Internet Protocol (VoIP) applications. The forehead accelerometer, affixed via a homemade headband, is a compact contact sensor that has been commercialized by Knowles. It is designed for use in high-noise environments and can be integrated into helmets for military and emergency service communications. Finally, the temple vibration sensor is an AKG product, a miniature vibration sensor originally designed for string instruments. In this study, this sensor has been diverted from its original purpose to capture bone vibrations by placing it in an adjustable 3D-printed enclosure adapted to the participants' zygomatic bones.\\

In addition to sensor specifications, Table \ref{tab:specifications} presents two signal-to-noise ratio (SNR) measures for each sensor. The first, signal-to-self-noise, is computed as the difference between the $L_{50}^{silence}$ and $L_{50}^{speech}$ values, which will be defined in Section \ref{subsec:post-processing}, across the entire \texttt{speechless-clean} split of the dataset, as defined in Section \ref{subsec:recording}. Notably, all sensors exhibit a high signal-to-self-noise ratio, except for the temple vibration sensor, which is primarily designed to capture much higher amplitude signals. The second measure is the signal-to-external-noise ratio, calculated under an ambient noise level of 90 dB. Since the reference signal is unavailable in this case, we derived this ratio by decomposing it into the signal-to-self-noise minus the external-noise-to-self-noise ratio. This was estimated using the \texttt{speechless-noisy} and \texttt{speechless-clean} subsets, as defined in Section \ref{subsec:recording}. This decomposition assumes that speakers produce speech at the same loudness in both quiet and noisy environments. However, in practice, the Lombard effect causes speakers to increase their vocal effort in noisy conditions, meaning the reported signal-to-external-noise ratios represent a lower bound on the actual values. Despite this approximation, the signal-to-external-noise ratio remains a direct indicator of each sensor’s noise resilience. Our results show that sensors directly capturing skin vibrations exhibit the highest noise resilience. However, these sensors also tend to have the smallest bandwidth, highlighting the trade-off between noise robustness and frequency range.

\subsection{Textual data}
\label{subsec:external_textual_data}

The text utilized for participant readings originates from the French Wikipedia subset of Common Voice \cite{ardila2019common}. This textual data was further filtered to produce a simplified dataset with a minimum number of textual and phonemic symbols and to minimize pronunciation uncertainty. To create this simplified dataset, we applied textual filters that excluded most proper names, retaining only common first names, French town or region names, and country names. Moreover, utterances containing numbers, Greek letters, mathematical symbols, or syntactic errors were carefully removed, resulting in a more straightforward dataset that is suitable for accurate pronunciation analysis. Once the unwanted sentences were removed, we applied common text normalization operations, such as lowercasing text and removing punctuation except for apostrophes. From this normalized text, we also generated the corresponding phonetic transcription using the International Phonetic Alphabet (IPA), which contains 33 phonemes in French. This operation was done with the text phonemizer of the CoML lab \cite{Bernard2021}. For some words, the phonemizer produced some language switches that we wanted to avoid, to preserve the minimal French phonetic alphabet. A phase of manual phonetic transcription was then necessary for these unrecognized words, such as "\textit{algorithme}".

\subsection{External ambiant noise data}
\label{subsec:external_audio_data}

The Vibravox contains 4 subsets: \texttt{speech-clean}, \texttt{speech-noisy}, \texttt{speechless-clean} and \texttt{speechless-noisy}. For the \texttt{speech-noisy} and \texttt{speechless-noisy} subsets, audio recordings were obtained from sensors worn by participants (either speaking or remaining silent) during the broadcast of 3D spatialized ambient noise. For all these ambient noise samples, the spatialization process was carried out using the \verb+ambitools+ library\footnote{available at \url{https://github.com/sekisushai/ambitools}} \cite{lecomte2015real} and rendered by a spherical array with a radius of 1.07 meters, made up of 56 loudspeakers placed around the participants \cite{lecomte2016fifty}. A significant proportion of the noise excerpts were sourced from the Audio Set \cite{gemmeke2017audio}, with a particular focus on noise classes relevant to BCMs applications, such as those encountered in battlefield and industrial environments. The Audio Set noise samples were spatialized using a plane wave model and from random fixed positions uniformly distributed on the sphere. In addition, in-house recordings of concerts and street events in 3\textsuperscript{rd} and 5\textsuperscript{th} order ambisonics were also rendered around the participants.

\subsection{Recording protocol}
\label{subsec:recording}
The recording process consists of four steps that are performed consecutively for each participant in the order listed below. Each of these steps corresponds to a subset of the Vibravox dataset.

\begin{itemize}

 \item \texttt{speech-clean}: The participant reads sentences sourced from the French Wikipedia for a duration of 15 minutes. Each utterance generates a new recording, and the transcriptions are retained. The recordings from this step populate the Vibravox subset that contains the most data for training.

 \item \texttt{speechless-noisy}: For a period of 2 minutes and 24 seconds, the subject is required to remain silent yet free to move, swallow, and breathe naturally in a noisy environment created from the AudioSet samples described in subsection \ref{subsec:external_audio_data}. Those samples are played in a spatialization sphere equipped with 56 loudspeakers surrounding the subject. The objective of this phase is to gather realistic background noises that will be combined with the \texttt{speech-clean} recordings to maintain a clean reference.

 \item \texttt{speechless-clean}: The procedure is repeated for 54 seconds in complete silence to record solely physiological and audio sensor noises. These samples can be valuable for tasks such as heart rate tracking or simply analyzing the noise properties of the various sensors. It could also be conveniently used to generate synthetic datasets with realistic physiological (and sensor-inherent) noise captured by body-conduction audio sensors.

\item \texttt{speech-noisy}: The final phase (54 seconds) will serve to test the different systems (speech enhancement, automatic speech recognition, speaker verification, ...) that will be developed based on the recordings from the first three subsets. This real-world testing will provide valuable insights into the performance and effectiveness of these systems in practical scenarios. In this phase, the noise samples replayed in the 3D spatializer are sourced from in-house ambisonic recordings of concerts and street events.

\end{itemize}

\begin{figure*}[ht!]
         \centering
         \subfloat[Extracting indicators]{\includegraphics[width=0.93\linewidth]{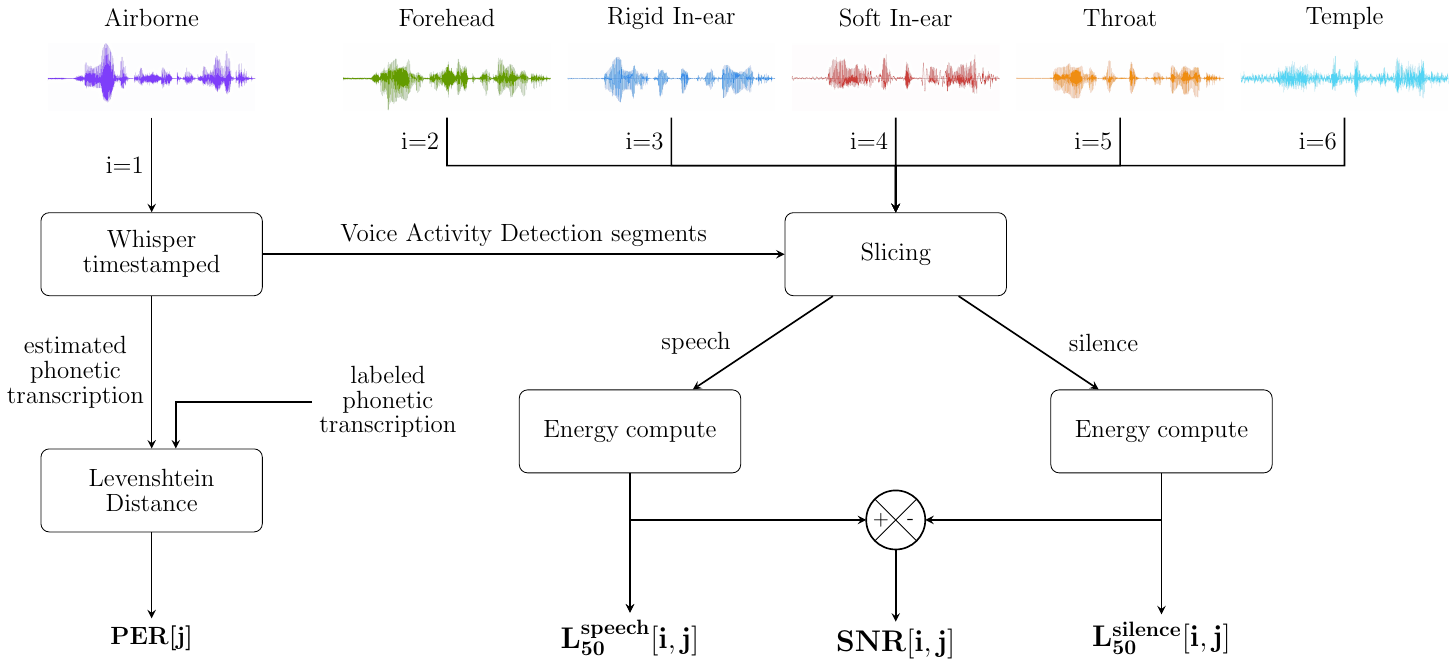}} \\
         \subfloat[Filtering process]{\includegraphics[width=0.93\linewidth]{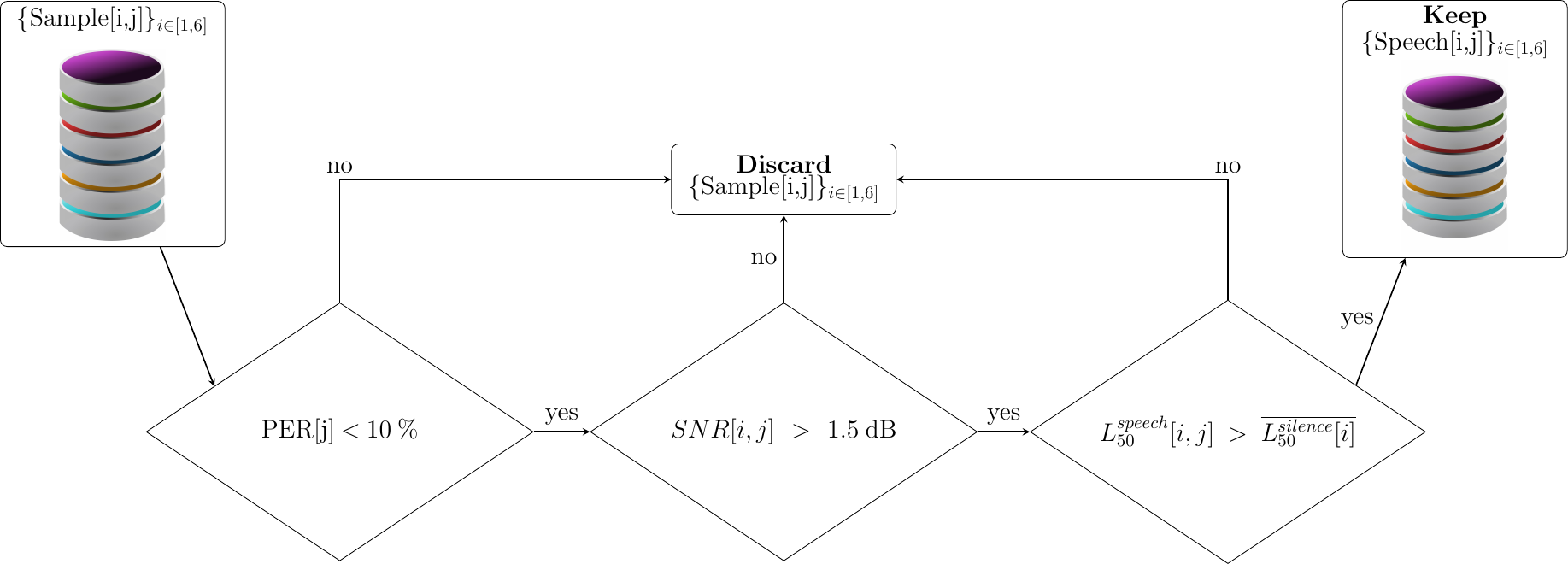}} \\
         \caption{Post-processing filtering process}
          \label{fig:tikz}
\end{figure*}

\begin{table*}[h]
\centering
\caption{Splits and subsets data summary after post-processing which is the publicly released version}
\begin{tabular}{@{}cccccccccc@{}}
\toprule
Subset & Split & \makecell{Audio duration \\ (hours)} & Number of audio clips & \makecell{Number of Speakers \\ (Female/Male)} & \makecell{Median \\audio \\duration (s)} & \makecell{Max \\audio \\duration (s)} & \makecell{Min \\audio \\duration (s)} \\
\midrule
\multirow{3}{*}{\texttt{speech-clean}} & train & 6x26.34 & 6x20,981 & 77F/72M & 4.43 & 13.03 & 1.1 \\
 & validation & 6x3.11 & 6x2,523 & 9F/9M & 4.36 & 10.64 & 1.47 \\
 & test & 6x3.85 & 6x3,064 & 11F/10M & 4.44 & 10.27 & 1.38 \\
\midrule
\multirow{3}{*}{\texttt{speech-noisy}} & train & 6x1.57 & 6x1,220 & 77F/72M & 4.59 & 9.86 & 1.36 \\
 & validation & 6x0.17 & 6x132 & 9F/9M & 4.47 & 8.56 & 2.3 \\
 & test & 6x0.23 & 6x175 & 11F/10M & 4.7 & 7.67 & 1.85 \\
\midrule
\multirow{3}{*}{\texttt{speechless-clean}} & train & 6x2.24 & 6x149 & 77F/72M & 54.10 & 54.10 & 53.99 \\
 & validation & 6x0.27 & 6x18 & 9F/9M & 54.10 & 54.10 & 54.05 \\
 & test & 6x0.32 & 6x21 & 11F/10M & 54.10 & 54.10 & 54.10 \\
\midrule
\multirow{3}{*}{\texttt{speechless-noisy}} & train & 6x5.96 & 6x149 & 77F/72M & 144.03 & 144.17 & 143.84 \\
 & validation & 6x0.72 & 6x18 & 9F/9M & 144.03 & 144.05 & 143.94 \\
 & test & 6x0.84 & 6x21 & 11F/10M & 144.03 & 144.05 & 144.03 \\
\midrule
\textbf{Total} & & \textbf{6x45.62} & \textbf{6x28,471} & \textbf{97F/91M} & & & & & \\
\bottomrule
\end{tabular}
\label{tab:audio_data_summary}
\end{table*}

\subsection{Post processing}
\label{subsec:post-processing}


A total of $J$=30,568 sentences were initially collected from 200 participants in the \texttt{speech-clean} subset. However, a small number of audio clips had various shortcomings: although we carefully monitored and validated each recording during the experimental process, the workflow was not entirely foolproof. Mispronounced sentences, sensors shifting from their initial positions, occasional sensor malfunctions, or instances where a microphone was inadvertently left unequipped led to some data inconsistencies. In order to ensure that our dataset would allow for meaningful error analysis between sensors across multiple speech processing tasks, we applied a series of three automatic filters to remove utterances with obvious deficiencies. We retained only sentences with correct pronunciation and a sufficiently high SNR, based on predefined criteria outlined in Figure~\ref{fig:tikz}. However, in some cases, such as when a participant’s ear canal was too small for the rigid in-ear microphone to achieve proper acoustic sealing, these filters were not triggered.\\

The filtering process comprises two main stages. In the first stage, indicators for evaluating audio capture quality are extracted (Figure \ref{fig:tikz}(a)). In the second stage, these indicators are used to derive a set of filters (Figure \ref{fig:tikz}(b)). The whisper-timestamped model\footnote{available at \url{https://github.com/linto-ai/whisper-timestamped}} constitutes a key component of our indicator extraction process. This model is a speech-to-text engine based on Whisper \cite{radford2022whisper} that applies dynamic time warping \cite{JSSv031i07} to cross-attention weights. By aligning the transcription with the raw waveform, accurate prediction of word timestamps in speech segments can be used to identify sequences containing vocalizations. The indicators we are using are listed below:

\begin{itemize}

\item $\mathrm{PER}[j]$: Phoneme Error Rate is computed for each sample $j \in [1,J]$ using the Levenshtein distance between the phonetic labeled transcription and the phonetic transcription derived from Whisper's output. Note that we used the exact same phonemization procedure as described in subsection \ref{subsec:external_audio_data}.

  \item $L_{50}^{silence}[i,j]$: In acoustics, percentile levels are statistical measures used to describe the distribution of sound levels over a specified period \cite{crocker2007rating}. These levels indicate the sound level that is exceeded for a certain percentage of the time. In this case, $L_{50}^{silence}[i,j]$ represents the sound level (in dB) exceeded 50\% of the time during non-voiced segments of the sample $j$ recorded on sensor $i$. The "silence" segments are obtained by running the whisper-timestamped model on the headset microphone signals, which are considered as the reference in quiet conditions. The fluctuating sound levels for calculating $L_{50}$ are determined using windows of 4092 samples at the original 48 kHz sampling rate. Using $L_{50}^{silence}[i,j]$, $\overline{L_{50}^{silence}[i]}$ is also derived, which represents the mean $L_{50}$ level for silence recorded by each sensor $i$.

\item $L_{50}^{speech}[i,j]$: The sound level exceeded 50\% of the time (in dB) during voiced segments of the sample $j$ recorded on sensor $i$. The same processing as $L_{50}^{silence}[i,j]$ was used.

\end{itemize}

The aforementioned indicators are then employed in three filters. The first filter discards any audio samples where $\mathrm{PER}[j]~>~10\%$. This addresses potential discrepancies between the labeled transcription and actual pronunciation, ensuring high-quality labels for the speech-to-phonemes task. The second filter is employed to verify that the sensor is functioning correctly, specifically by examining the ratio between speech and silence energy levels on a given sample. This ratio, denoted by $ \mathrm{SNR}[i,j] = L_{50}^{speech}[i,j] - L_{50}^{silence}[i,j] > 1.5dB $, is indicative of recordings with low vocal energy or those that have been affected by sensor malfunction. The last filter is designed to detect any potential sensitivity drift on the sensors. Such drift could be caused by a bug in the electronics or mechanical blockage of the transducer. To this end, the filter checks that $L_{50}^{speech}[i,j] > \overline{L_{50}^{silence}[i]}$. Only \texttt{speech-clean} samples that pass the three filters are added to the Vibravox dataset. We systematically added all audio recordings for the three other subsets (\texttt{speech-noisy}, \texttt{speechless-clean}, \texttt{speechless-noisy}) if the corresponding \texttt{speech-clean} subset of the participant is not empty. Out of 200 participants, 188 were retained, with a total of 28,471 sentences. Given that this post-processing filtering excluded a non-negligible amount of data (6~$\%$ of participants and 6.8~$\%$ of pronounced sentences), we emphasize that its primary objective is to ensure rigorous analysis and fair comparison between microphones in subsequent speech processing tasks, rather than to maximize overall performance. The dataset is then split by speaker, with 80\% of male and female speakers in the training set, 10\% in the development/validation set, and 10\% in the test set. This ensures that all utterances from a given speaker are grouped together, preventing any speaker overlap between splits. Once assigned to a split, a participant remains in the same split across all subsets. All the details relative to the final version of the Vibravox dataset are presented in Table \ref{tab:audio_data_summary}.

Note that if a sample $j$ passes all filters, it is not immediately added to the dataset. Instead, the timestamps generated by Voice Activity Detection (VAD) are extended by 0.6 seconds on both sides. In rare cases where the adjusted start or end timestamps exceed the original audio boundaries, the audio is trimmed by 15 ms on the affected side. This approach helps eliminate unwanted artifacts, such as mouse clicks at segment edges, while ensuring more reliable capture of vocal segments without discarding valid speech.

\subsection{Signal processing analysis}

\begin{figure}[H]
  \centering
  \centerline{\includegraphics[width=0.98 \linewidth]{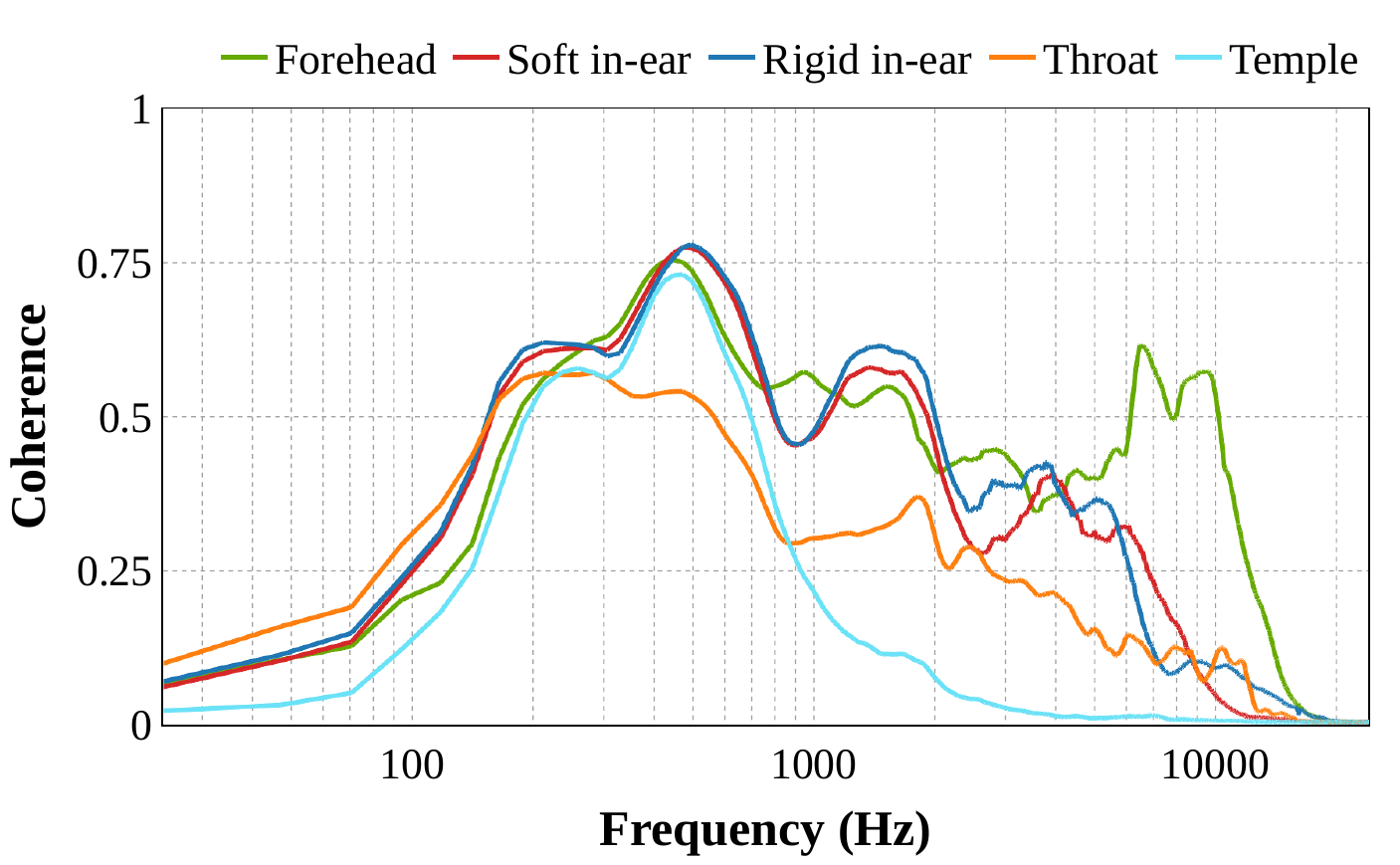}}
\caption{Coherence functions of body-conducted sensors in quiet conditions}
\label{fig:coherence}
\end{figure}

The \texttt{speech-clean} subset of the Vibravox dataset provides an ideal resource for investigating the differences between various audio sensors. Figure \ref{fig:coherence} illustrates the coherence functions of all BCMs, based on the source-filter model, where the headset microphone serves as the input and the corresponding BCMs serve as the output. These coherence functions were averaged on the entire filtered dataset and computed using 2,048 frequency bins on the raw signals recorded at 48 kHz.


Each sensor reveals a distinct bandwidth size that the figure legend follows from left to right in decreasing order. The temple sensor exhibits the smallest bandwidth, barely recording any speech signal above 1500 Hz. The laryngophone follows, with coherence dropping below 0.25 at 2000 Hz. The soft and rigid in-ear sensors exhibit larger bandwidths yet display clear antiresonance frequencies. The soft in-ear bandwidth is slightly larger than its rigid in-ear counterpart, which may be due to a less effective acoustic seal. Finally, the forehead accelerometer offers the largest bandwidth but does not effectively filter out external noise pollution. As discussed in Section \ref{sec:tasks}, this bandwidth is closely correlated with the performance of deep learning models on speech intelligibility, quality, speaker verification, and speech-to-phonemes.

Beyond bandwidth analysis, we can also observe that the shape of these coherences is not always flat, depending on the sensor placement. We cannot draw clear conclusions here because we also changed the technology at each placement, but some works like \cite{mcbride2011effect} have already explored the question of the best body-conducted microphone location. Their findings show that the forehead and temple are the best spots to capture speech with the highest intelligibility and sound quality, while the throat is among the worst positions.

\section{Using Vibravox}
\label{sec:tasks}

The Vibravox dataset offers considerable potential across various applications, particularly in improving communication in noisy environments. The \texttt{speech-noisy} subset is of particular interest for showcasing the noise-resilient capabilities of BCMs. However, this subset primarily serves testing purposes, as it lacks a clean reference under such conditions. As a consequence, the majority of the dataset is contained within the \texttt{speech-clean} subset, where the airborne microphone serves as the reference. In order to simulate real-world noisy scenarios while retaining a clean reference, data augmentation techniques can be applied by adding audio segments of the \texttt{speechless-noisy} subset with BCM utterances. Furthermore, the Vibravox dataset displays potential for applications such as developing voice commands and monitoring vital signs, including heart rate, particularly when utilizing the \texttt{speechless-clean} subset. In the following, we propose three classic tasks in automatic speech processing to investigate the distinctive characteristics of BCMs, establishing baselines for each task.

\subsection{Speech Enhancement}
\label{subsec:se}

The high impedance of BCMs enables them to focus on the wearer's voice by rejecting ambient noise, but this also affects their ability to capture mid and high frequency components due to the intrinsic low-pass characteristics of the biological pathway of vibrations between the vocal cords and the body-conducted audio sensors. The absence of these frequencies in captured speech directly impacts the perceived quality, intelligibility, and distinctiveness. This highlights the necessity for a post-capture enhancement algorithm.\\

\subsubsection{{Model}} We adopted the configurable EBEN \cite{hauret2023configurable} approach, enriched with recent trends in the audio field. This GAN \cite{goodfellow2014generative} architecture offers several advantages. In particular, the lightweight design of the 1D U-net-like generator is compatible with integration in mobile and wearable systems, where memory is limited and real-time is a priority. Additionally, the PQMF bank \cite{nguyen1994near} offers the advantage of manipulating frequency content without the need to consider phase or manage complex values, as it operates directly on waveform signals. It is important to note that the limitations of GANs, including the difficulty of reaching a Nash equilibrium, the lengthy training period, and the additional computational cost associated with the discriminators, do not affect the model once it has been deployed. \\

\subsubsection{{Architecture configuration}}

Separate EBEN models were trained for each sensor in the Vibravox dataset, using sensor-specific hyperparameters detailed in Table \ref{tab:stoi_and_mos_clean}. The original number of subbands was maintained at a value of $M=4$ for each sensor due to the significant variability in intra-sensor bandwidth among participants, rendering precise spectrogram slicing impractical. Furthermore, a smaller number of subbands allows for a more permissive design of the passband filters with a softer slope for the same resulting aliasing, thus shortening the PQMF kernel length, which is useful for reducing the algorithmic latency. The number of subbands provided to the model, denoted as $P$, was tailored to match the bandwidth of each sensor. For example, the temple vibration pickup captures signals within the 0-2 kHz range, while the forehead accelerometer exhibits good coherence with the reference microphone up to 8 kHz. Consequently, after resampling the signals to 16 kHz for training, only the first band was retained for the temple vibration pickup, while all bands were retained for the accelerometer. In addition, the last configurable parameter, $Q$, which represents the number of bands fed to the PQMF discriminator, has been modified to match $M$ at 4 for all models. It was observed that even when the audio sensors capture information within a specific frequency range -- including the corresponding subbands in the discriminator input -- it is beneficial for denoising minor artifacts present in these initial bands. In the case of speech captured in noisy conditions, even though body conduction sensors are robust to external acoustic pollution, some noise is still captured, providing an additional reason to focus discriminators work on every subband. This minor modification necessitated an adjustment to the number of channels in the PQMF discriminators in order to align with the divisibility constraints of the grouped convolution.\\

\begin{figure*}[ht!]
  \centering
  \centerline{\includegraphics[width=0.95 \linewidth]{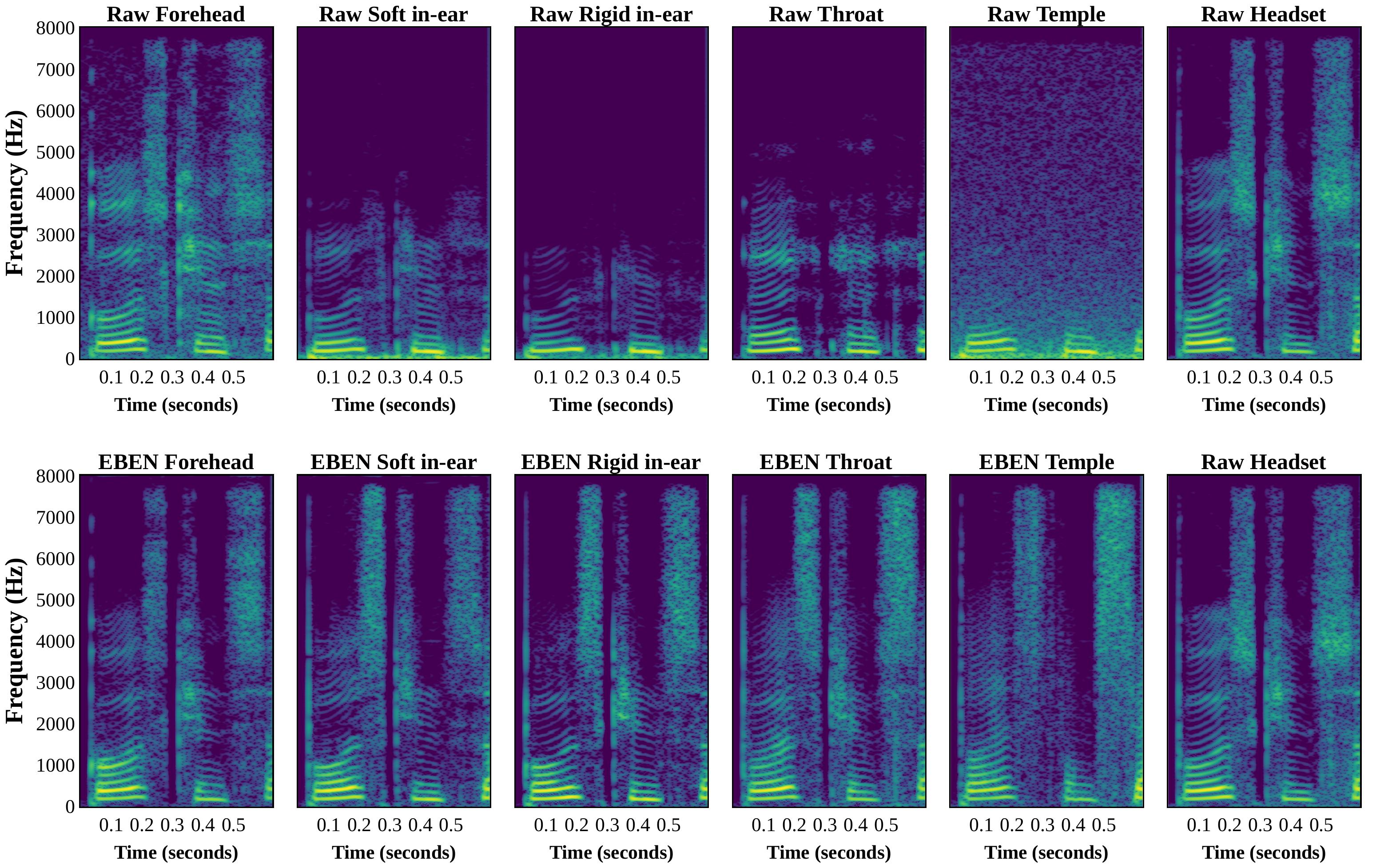}}
\caption{Spectrograms of signals recorded by the different Vibravox audio sensors in the \texttt{speech-clean} subset and their corresponding EBEN-enhanced version}
\label{fig:spec}
\end{figure*}

\subsubsection{{Training procedure}}
In comparison to the original implementation proposed in \cite{hauret2023configurable}, we have incorporated a spectral loss $\mathcal{L}_\mathcal{G}^{spec}$ into the generator objective. This modification proved to be effective in stabilizing the training process. This loss incorporates the same FFT, hop size, and window length parameters as described in \cite{yamamoto2020parallel}. Implementation is taken from \cite{steinmetz2020auraloss}. In contrast to some speech enhancement papers \cite{defossez2020real,subakan2021attention}, we did not include a reconstructive loss in the temporal domain due to the peculiarities of body-conduction speech. In fact, complex phase shifts occur : attempting to align the signals to the reference at the sample level degraded performance. We retained the original EBEN implementation's use of the other two adversarial and a feature matching losses, respectively noted $\mathcal{L}_\mathcal{G}^{adv}$ and $\mathcal{L}_\mathcal{G}^{feat}$. Also note that a slight adjustment to $\mathcal{L}_\mathcal{G}^{feat}$ has been done by normalizing it using the enhanced feature norm. The training strategy remained consistent across \texttt{speech-clean} and \texttt{speech-noisy} experiments, with a constant learning rate of $3e^{-4}$. To ensure reproducibility of experiments, the code has been made available on the GitHub repository \url{https://github.com/jhauret/vibravox}, which makes use of the following libraries \cite{paszke2019pytorch,falcon2019pytorch,Yadan2019Hydra,lhoest-etal-2021-datasets}.\\

\subsubsection{{Experiments in quiet conditions}}

For this initial set of experiments, the goal is to evaluate the performance of bandwidth extension in quiet conditions by training EBEN models on the \texttt{speech-clean} subset. A total of 500 epochs were required for convergence on the Vibravox dataset. To address the relatively small amount of data, light data augmentation was applied, including speed perturbation, pitch shifting, and time masking, all implemented in the Torchaudio library \cite{hwang2023torchaudio}. The correlation analysis presented in the configurable EBEN article \cite{hauret2023configurable} revealed that STOI \cite{taal2010short,jensen2016algorithm} and Noresqa-MOS (N-MOS) \cite{noresqamos} were the metrics with the highest correlation with MUSHRA studies in the context of speech enhancement on BCMs simulated data. Consequently, these two metrics were retained for inclusion in Table \ref{tab:stoi_and_mos_clean}, which presents the results of the bandwidth extension for each sensor in quiet conditions. The results reveal a clear relationship between the metrics value and the sensors' available bandwidth. Yet, both STOI and N-MOS performance appear to hit a glass ceiling, plateauing at 0.88 and 4.3, respectively. This saturation could stem from the challenge of extracting fine details from body-conducted signals. While striving to produce authentic samples, the generator only picks one plausible clean signal out of all the possible ones, typical of one-to-many problems \cite{tan2024regeneration}. The spectrograms in Figure~\ref{fig:spec} provide a visual representation of the processing.\\

\begin{table}[ht]
\caption{Testing of EBEN models and raw signals on \texttt{speech-clean}}
\vspace{-2mm}
\centering
\begin{tabular}{@{}lccc@{}}
\toprule
\textbf{Sensor} & \textbf{Configuration} & \textbf{STOI} & \textbf{Noresqa-MOS} \\
\midrule
Forehead       & Raw signal             & 0.731         & 3.760                \\
               & EBEN (M=4, P=4, Q=4)   & \textbf{0.855}         & \textbf{4.250}                \\
\midrule
Soft In-ear    & Raw signal             & 0.752         & 3.315                \\
               & EBEN (M=4, P=2, Q=4)   & \textbf{0.868}         & \textbf{4.331}                \\
\midrule
Rigid In-ear   & Raw signal             & 0.782         & 3.392                \\
               & EBEN (M=4, P=2, Q=4)   & \textbf{0.877}         & \textbf{4.285}                \\
\midrule
Throat         & Raw signal             & 0.677         & 3.097                \\
               & EBEN (M=4, P=2, Q=4)   & \textbf{0.834}         & \textbf{3.862}                \\
\midrule
Temple         & Raw signal             & 0.602         & 2.905                \\
               & EBEN (M=4, P=1, Q=4)   & \textbf{0.763}         & \textbf{3.632}                \\
\bottomrule
\end{tabular}
\label{tab:stoi_and_mos_clean}
\end{table}

\subsubsection{{Experiments in noisy conditions}}

After analyzing speech enhancement in quiet conditions, the next focus is a joint denoising and bandwidth extension task for recordings in noisy environments. To this end, the model is tested on the \texttt{speech-noisy} subset; however, this subset cannot be used for training due to the absence of a clean reference. For training, the \texttt{speechless-noisy} subset is leveraged and dynamically mixed with the \texttt{speech-clean} subset. Several configurations were tested, including retraining the EBEN models from scratch for 500 epochs or initializing from the previously trained \texttt{speech-clean} models, which converged after no more than 200 epochs for microphones with the largest domain gap. Additionally, \texttt{speech-clean} models were directly tested on \texttt{speech-noisy}. Hyperparameters of the training strategy were kept consistent with quiet condition experiments, with the same values for M, P, and Q across all microphones. \\

 Prior to testing on \texttt{speech-noisy}, the model performance was first evaluated on the validation and test splits of the synthetically mixed \texttt{speech-clean}+\texttt{speechless-noisy} data. When testing on \texttt{speech-noisy}, no significant differences were observed in the Noresqa-MOS scores, reflecting a realistic modeling of the training data. When using \texttt{speech-noisy} for evaluating the models, the STOI -- which is a reference-based metric -- was replaced with the reference-free TorchSquim-STOI \cite{kumar2023torchaudio}, which provides an estimate of the same quantity. Although direct comparison between STOI and TorchSquim-STOI is not possible, both metrics showed improvement during training, and the TorchSquim-STOI results remained consistent across both synthetic mixed and real noisy data. The final test results on \texttt{speech-noisy} are presented in Table~\ref{tab:stoi_and_mos_noisy}.

\begin{table}[ht]
\caption{Testing of EBEN models and raw signals on \texttt{speech-noisy}}
\vspace{-2mm}
\centering
\resizebox{\columnwidth}{!}{
\begin{tabular}{@{}lccc@{}}
\toprule
\textbf{Sensor} & \textbf{Initialization} & \textbf{Squim-STOI} & \textbf{Noresqa-MOS} \\
\midrule
Forehead       & Raw signal                & 0.901 & 3.85  \\
               & Tested from pretrained$^{*}$   & 0.949 & 4.08  \\
\addlinespace[2pt]
\cline{2-4} 
\addlinespace[3pt]
               & Trained$^{\dagger}$ from scratch~~   & 0.949 & 4.14  \\
               & Trained$^{\dagger}$ from pretrained$^{*}$   & \textbf{0.971} & \textbf{4.20}  \\
\midrule
Soft In-ear    & Raw signal                & 0.813 & 3.38  \\
               & Tested from pretrained$^{*}$   & 0.865 & 3.71  \\
\addlinespace[2pt]
\cline{2-4} 
\addlinespace[3pt]
               & Trained$^{\dagger}$ from scratch~~   & 0.911 & 3.63  \\
               & Trained$^{\dagger}$ from pretrained$^{*}$   & \textbf{0.917} & \textbf{3.87}  \\
\midrule
Rigid In-ear   & Raw signal                & 0.751 & 3.47  \\
               & Tested from pretrained$^{*}$   & 0.812 & 3.73  \\
\addlinespace[2pt]
\cline{2-4} 
\addlinespace[3pt]
               & Trained$^{\dagger}$ from scratch~~   & \textbf{0.876} & 3.51  \\
               & Trained$^{\dagger}$ from pretrained$^{*}$   & 0.873 & \textbf{3.80}  \\
\midrule
Throat         & Raw signal                & 0.942 & 3.71   \\
               & Tested from pretrained$^{*}$   & 0.969 & \textbf{4.05}  \\
\addlinespace[2pt]
\cline{2-4} 
\addlinespace[3pt]
               & Trained$^{\dagger}$ from scratch~~   & \textbf{0.978} & 3.87  \\
               & Trained$^{\dagger}$ from pretrained$^{*}$   & 0.971 & 3.98  \\
\midrule
Temple         & Raw signal                & 0.848 & 3.46  \\
               & Tested from pretrained$^{*}$   & 0.954 & \textbf{3.77}  \\
\addlinespace[2pt]
\cline{2-4} 
\addlinespace[3pt]
               & Trained$^{\dagger}$ from scratch~~   & \textbf{0.966} & \textbf{3.77}  \\
               & Trained$^{\dagger}$ from pretrained$^{*}$   & 0.961 & 3.63  \\
\bottomrule
\addlinespace[3pt]
\multicolumn{4}{l}{\textit{$^{*}$ on the \texttt{speech-clean} dataset}} \\
 \multicolumn{4}{l}{$^{\dagger}$ \textit{on the synthetically mixed \texttt{speech-clean}+\texttt{speechless-noisy} dataset}} \\
\end{tabular}
}
\label{tab:stoi_and_mos_noisy}
\end{table}

The results reveal that microphones with the lowest signal-to-external-noise ratios benefited the most from the retrained strategies. Some microphones, such as the temple vibration pickup, were so insensitive to noise that the non-retrained \texttt{speech-clean} models outperformed them. The Noresqa-MOS scores for the raw signals also increased compared to the \texttt{speech-clean} subset. However, these Noresqa-MOS values cannot be directly compared, as in the quiet conditions, the metric was computed using matched signals, whereas in the noisy conditions, a random headset signal from the \texttt{speech-clean} was used for the calculation. \\

\begin{figure*}[ht!]
  \centering
  \centerline{\includegraphics[width=0.95 \linewidth]{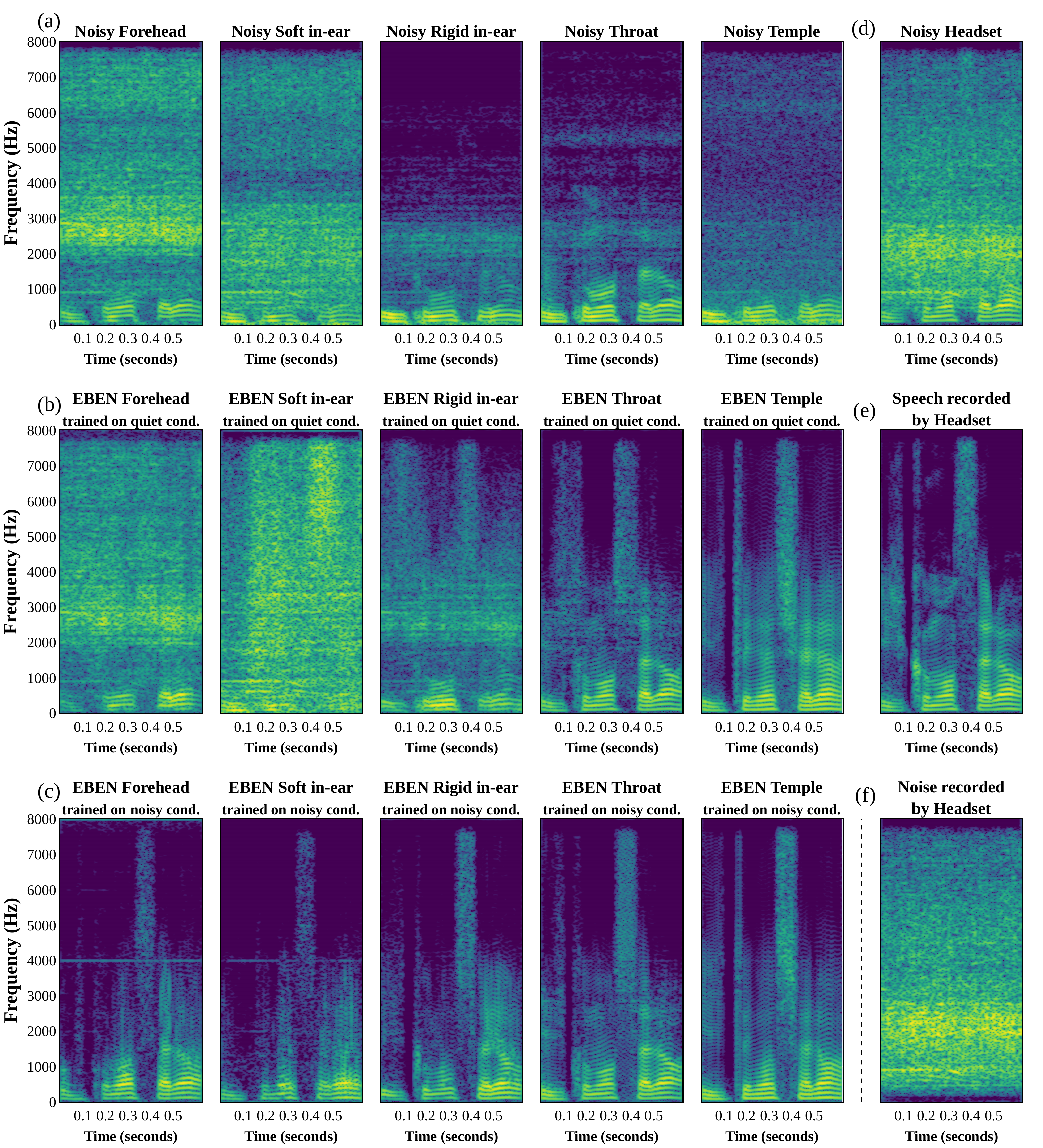}}
\caption{Synchronized spectrograms of multiple signals for the speech enhancement task in noisy conditions.
(a) \& (d): Synthetically mixed samples, speech comes from \texttt{speech-clean} and noise comes from \texttt{speechless-noisy} respective microphones
 - (b): Corresponding enhanced samples performed by EBEN models trained on the \texttt{speech-clean} subset
 - (c): Corresponding enhanced samples performed by EBEN models trained on the synthetically mixed \texttt{speech-clean} and \texttt{speechless-noisy} subsets
 - (e): Clean speech sample recorded by the headset microphone in the \texttt{speech-clean} subset before mixing
 - (f): Noise sample recorded by the headset microphone in the \texttt{speechless-noisy} subset before mixing}
\label{fig:spec-noisy}
\end{figure*}

As with the experiments conducted in quiet conditions, several spectrograms are presented in Figure \ref{fig:spec-noisy}. Synthetically mixed data were selected to visually assess the effect of the models, as both the noise and the clean target signals remain available. This figure highlights several noteworthy observations. Notably, the excessive noise detected in the soft in-ear microphone suggests an insertion defect for this particular sample of \texttt{speechless-noisy}, whereas the rigid microphone appears to provide better acoustic sealing. However, such defects occur just as often in both microphone types and are absent in most cases. Additionally, EBEN models trained for bandwidth extension in quiet conditions exhibit limited noise suppression. In contrast, as shown in line (c), noise-robust EBEN models effectively suppress noise, with the difference particularly pronounced for forehead and in-ear microphones. For throat and temple microphones, which are inherently more resistant to noise, the domain gap between quiet and noisy conditions remains small, enabling the \texttt{speech-clean} EBEN models to maintain strong performance. Achieving comparable denoising performance from the noisy headset using conventional denoising networks would be highly challenging.

\vspace{-1.5mm}

\subsection{Speech-to-Phonemes}
\label{subsec:stt}

Automatic Speech Recognition (ASR) systems serve as crucial facilitators of seamless human-computer interaction, allowing users to dictate text, control devices, and access information through spoken language. Proficiency in this realm with BCMs can confer significant advantages, particularly in military operations or industrial settings.\\

\subsubsection{{Model}} ASR is often divided into two models as in \cite{baevski2020wav2vec}: an acoustic model that infers the tokens from the raw waveform and a language model that is used to refine the probabilities given by the acoustic model with a beam search algorithm. However, our focus primarily revolves around investigating the acoustic characteristics of various sensors. Hence, we have opted to exclude the language model and instead predict phonemes directly with greedy decoding of the acoustic model. The tokens predicted by the linear head of our model correspond to the 33 phonemes of the French phonetic alphabet. We used a medium wav2vec2.0 \cite{baevski2020wav2vec} as our acoustic model, pretrained on the multilingual VoxPopuli speech corpus \cite{wang2021voxpopuli}.\\

\begin{table*}[ht!]
\caption{Top five editops count for in-distribution testing of audio phonemizers on raw speech signals in \texttt{speech-clean}}
\vspace{-2mm}
\centering
\begin{tabular}{@{}ccccccc@{}}
\toprule
\textbf{Sensor} & \textbf{1\textsuperscript{st} editop} & \textbf{2\textsuperscript{nd} editop}  & \textbf{3\textsuperscript{rd} editop} & \textbf{4\textsuperscript{th} editop}  & \textbf{5\textsuperscript{th} editop} \\
\midrule
\textbf{Headset Microphone}  & [o] $\rightarrow$ [\textipa{\textopeno}] (98) & [e] $\rightarrow$ [\textipa{\textepsilon}] (75) & [\textipa{\textscripta}] $\rightarrow$ [\textipa{\textopeno}] (64) & [\textipa{\textopeno}] $\rightarrow$ [o] (64) & [$\emptyset$] $\rightarrow$ [l] (52) \\
\textbf{Forehead Accelerometer}  & [$\emptyset$] $\rightarrow$ [l] (124) & [\textipa{\textopeno}] $\rightarrow$ [o] (106) & [\textipa{\textepsilon}] $\rightarrow$ [e] (92) & [o] $\rightarrow$ [\textipa{\textopeno}] (80) & [$\emptyset$] $\rightarrow$ [\textipa{\textepsilon}] (74) \\
\textbf{Soft In-ear Microphone}  & [\textipa{\textepsilon}] $\rightarrow$ [e] (127) & [\textipa{\textopeno}] $\rightarrow$ [o] (115) & [$\emptyset$] $\rightarrow$ [l] (93) & [\textipa{\textscripta}] $\rightarrow$ [\textipa{\textopeno}] (68) & [$\emptyset$] $\rightarrow$ [\textipa{\textepsilon}] (66) \\
\textbf{Rigid In-ear Microphone} & [\textipa{\textepsilon}] $\rightarrow$ [e] (122) & [$\emptyset$] $\rightarrow$ [l] (103) & [\textipa{\textopeno}] $\rightarrow$ [o] (97) & [o] $\rightarrow$ [\textipa{\textopeno}] (93) & [$\emptyset$] $\rightarrow$ [j] (84) \\
\textbf{Throat Microphone}  & [$\emptyset$] $\rightarrow$ [l] (160) & [$\emptyset$] $\rightarrow$ [\textipa{\textepsilon}] (148) & [$\emptyset$] $\rightarrow$ [i] (129) & [\textipa{\textopeno}] $\rightarrow$ [o] (128) & [$\emptyset$] $\rightarrow$ [\textipa{\textinvscr}] (107) \\
\textbf{Temple Vibration Pickup} & [$\emptyset$] $\rightarrow$ [s] (294) & [$\emptyset$] $\rightarrow$ [l] (290) & [$\emptyset$] $\rightarrow$ [t] (281) & [$\emptyset$] $\rightarrow$ [\textipa{\textinvscr}] (281) & [$\emptyset$] $\rightarrow$ [i] (229) \\
\bottomrule
\end{tabular}
\label{tab:phonetic_confusions}
\end{table*}

\subsubsection{{Experiments in quiet conditions}}
\begin{figure}
  \centering
  \includegraphics[width=0.98 \linewidth]{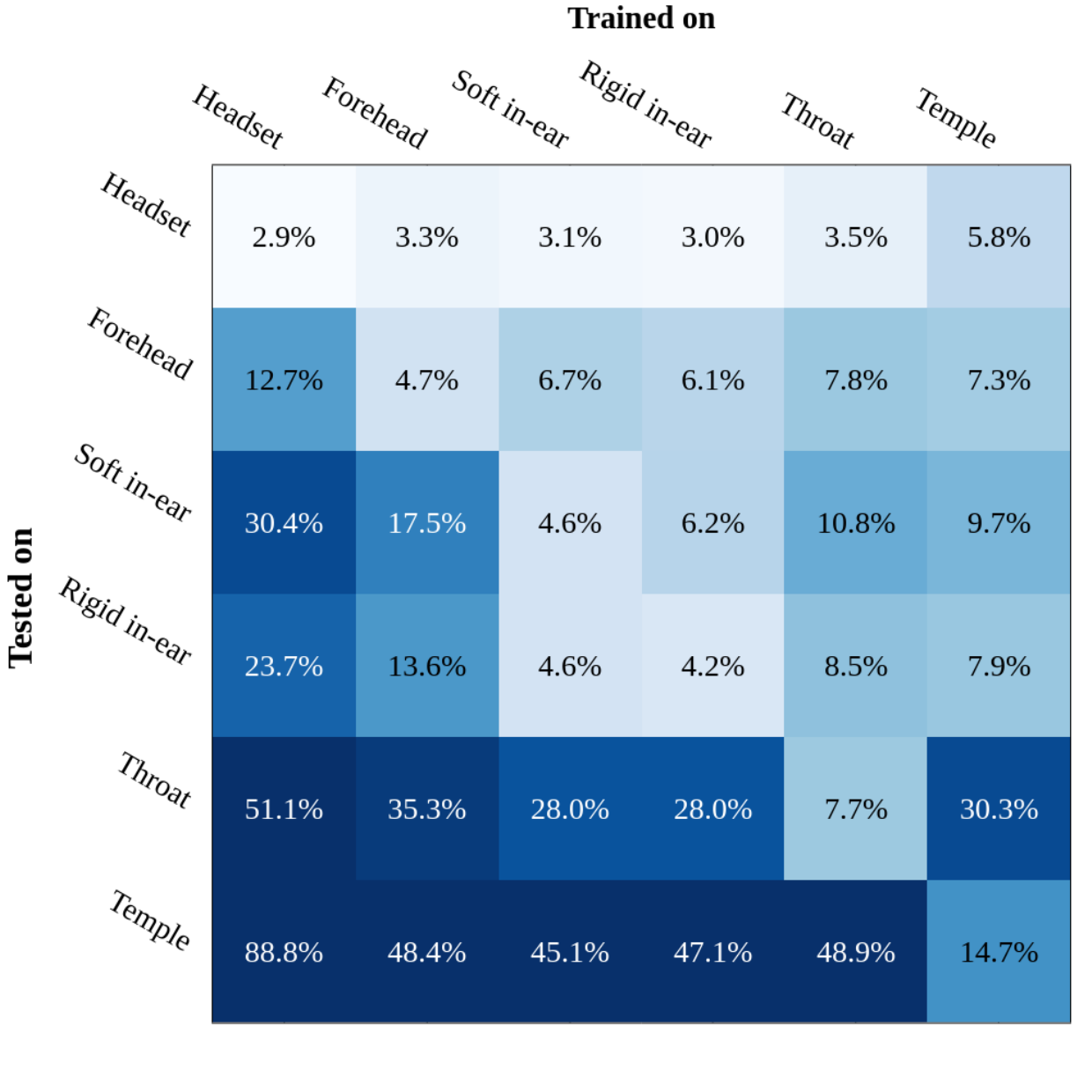}
  \caption{PER of the finetuned audio phonemizers. Models are trained and tested on the \texttt{speech-clean} subset}
  \label{img:performance_matrix_speech_clean}
\end{figure}

For this first set of experiments, we finetuned the wav2vec2.0 models with a constant learning rate of $1e^{-5}$ on the \texttt{speech-clean} subset of the Vibravox dataset for 10 epochs, minimizing the CTC loss \cite{graves2006connectionist}. We then tested each fine-tuned model not only on its corresponding test set but also on every other audio sensor, with the PER (Phoneme Error Rate) results shown in Figure \ref{img:performance_matrix_speech_clean}. As expected, the matrix shows dominance along the diagonal, which corresponds to in-distribution tests. From this diagonal, we infer that the temple vibration pickup and the throat microphone pose the greatest challenge for ASR in quiet conditions, given their limited bandwidth. A comparison between models trained on airborne and temple vibration sensors reveals a significant trend: the temple vibration pickup demonstrates enhanced robustness when tested on other sensor data. This trend can be attributed to the model's reliance primarily on low-frequency information, which is present on all sensors. Regarding the in-ear microphones, there is a possibility that a single model can be developed in future studies, since they are highly similar and exhibit minimal performance loss when cross-tested. At last, the performance of the forehead accelerometer is approaching that of the in-ear microphone, despite the larger bandwidth. This may be attributed to the considerable variance in the coherence function and the lower signal-to-self-noise ratio.\\

To gain a detailed understanding of the acoustic properties of each sensor, we also identified the top five phonetic confusions of all phonemizers when tested on their raw signals training distribution. Table \ref{tab:phonetic_confusions} offers valuable insights into the challenges inherent to each sensor placement and technology for the identification of phonemes due to the absence of high-frequency content. This table was obtained by counting the most frequently used editing operation to go from the model's decoded output to the labeled phonetic transcription when computing the Levenshtein distance \cite{levenshtein1966binary}. It should be noted that only in-word phonetic confusions were considered to avoid including the suprasegmental phenomenon [{\footnotesize \rotatebox{90}{(}}] (\textit{i.e.} "liaison" in French), which poses difficulties for many native speakers during reading and which is not present in our labeled phonetic transcription. This phenomenon caused all models to either omit or add [z] (\textit{i.e.} [$\emptyset$] $\rightarrow$ [z] and [z] $\rightarrow$ [$\emptyset$]) as their top errors. Once this issue has been disregarded, Table \ref{tab:phonetic_confusions} reveals that the model trained on airborne microphone signals encounters difficulties in distinguishing the similar phoneme confusion we encounter in French, such as [e]/[\textipa{\textepsilon}] and [\textipa{\textscripta}/][o]/[\textipa{\textopeno}], which are all voiced sounds. Similar errors have been observed in other audio sensors with a sufficiently large bandwidth, albeit with varying extents. Of particular interest is the observation that the temple vibration pickup, which has a narrow bandwidth, most frequently fails to capture the voiceless alveolar and plosive fricatives [s]/[t] that rely on substantial high-frequency support. The wideband support [\textipa{\textinvscr}] is also left out to a significant extent for the temple vibration pickup and throat microphone. For the temple sensor, this could be due to the similarity of this phoneme with the sensor's self-noise. For the throat microphone, we do not have a clear explanation, especially since this sensor placement seems to be optimal for capturing this uvular fricative. \\

\subsubsection{{Effect of EBEN enhancement in quiet conditions}}
\begin{figure}[h!]
  \centering
  \includegraphics[width=0.98 \linewidth]{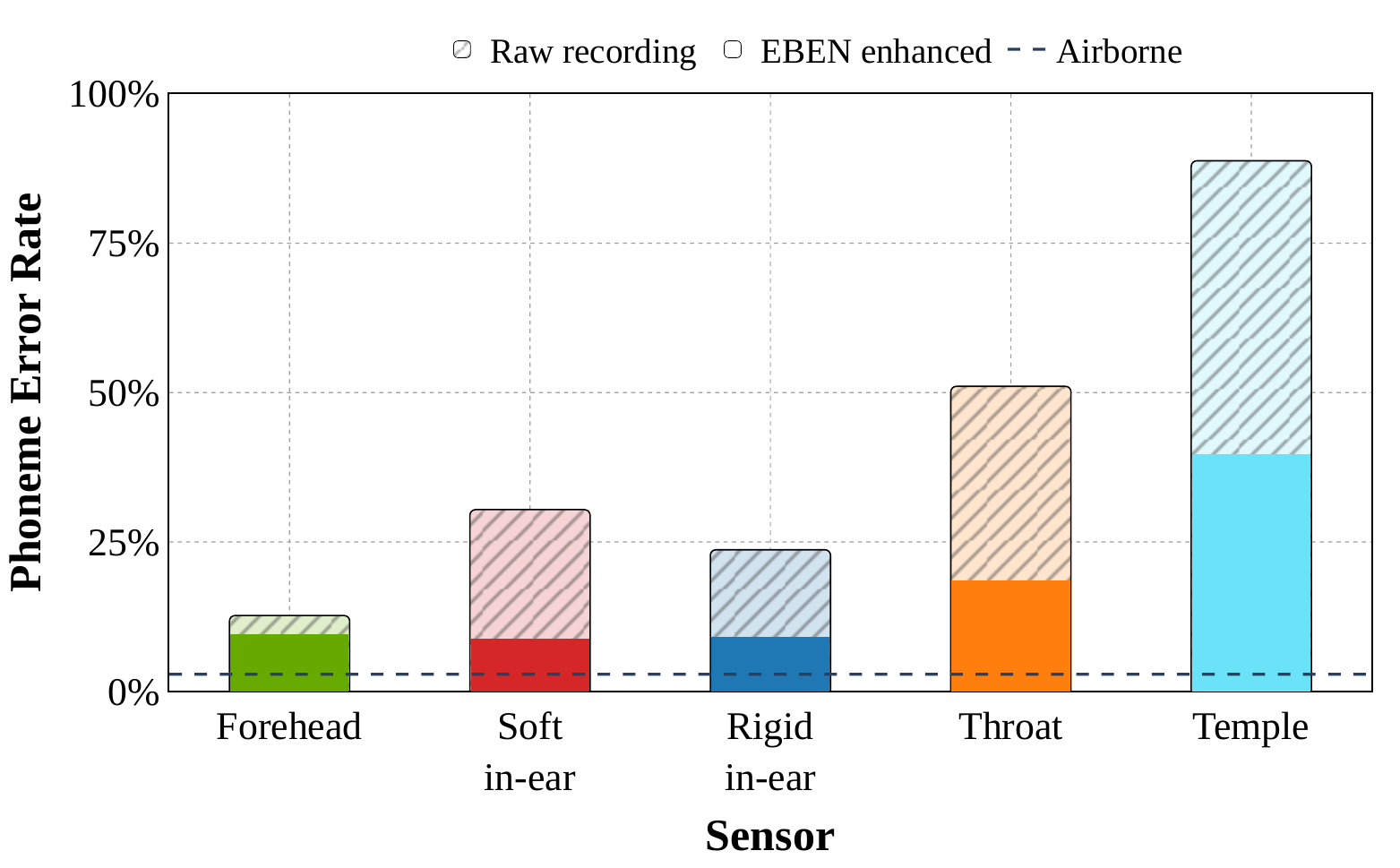}
  \caption{PER on \texttt{speech-clean} of raw and EBEN-enhanced test recordings phonemized by the headset phonemizer}
  \label{img:per_comparison_histogram}
\end{figure}

A supplementary analysis of the PER has been performed in order to quantify the effect of the bandwidth enhancement from EBEN models trained on \texttt{speech-clean} presented in Subsection \ref{subsec:se} for ASR. This comparison was conducted employing solely the headset phonemizer. The results presented as a bar graph in Figure \ref{img:per_comparison_histogram} demonstrate that all sensors exhibited a strong improvement in PER thanks to the enhancement obtained with EBEN models. This is a promising outcome, since contemporary speech enhancement systems, when used as a pre-processing layer in ASR systems, continue to face challenges in enhancing performance due to the production of artifacts \cite{iwamoto2022bad}. The aforementioned findings thus corroborate the hypothesis that speech intelligibility is enhanced by EBEN models. However, it is essential to note that the direct parallel between the headset phonemizer and the human ear cannot be taken for granted. Subjective testing could be considered to provide an additional layer of validation. \\

\subsubsection{{Experiments in noisy conditions}}
To complete the analysis of this task, a new set of audio phonemizers was trained on the \texttt{speech-noisy} training set, using the raw, unprocessed samples. Unlike for the speech enhancement task, where a clean reference is unavailable in \texttt{speech-noisy}, labels (phonemized transcriptions) remain available as reference data in noisy conditions. Several training strategies were explored, including retraining from the VoxPopuli checkpoint and fine-tuning the previously trained \texttt{speech-clean} models with an even smaller learning rate of $1e^{-6}$. The latter approach demonstrated a clear advantage. Additionally, given that the \texttt{speech-noisy} training set is only 1 hour and 34 minutes in length, an aggressive data augmentation strategy was necessary to prevent overfitting. \\

\begin{figure}[ht!]
  \centering
  \includegraphics[width=0.98 \linewidth]{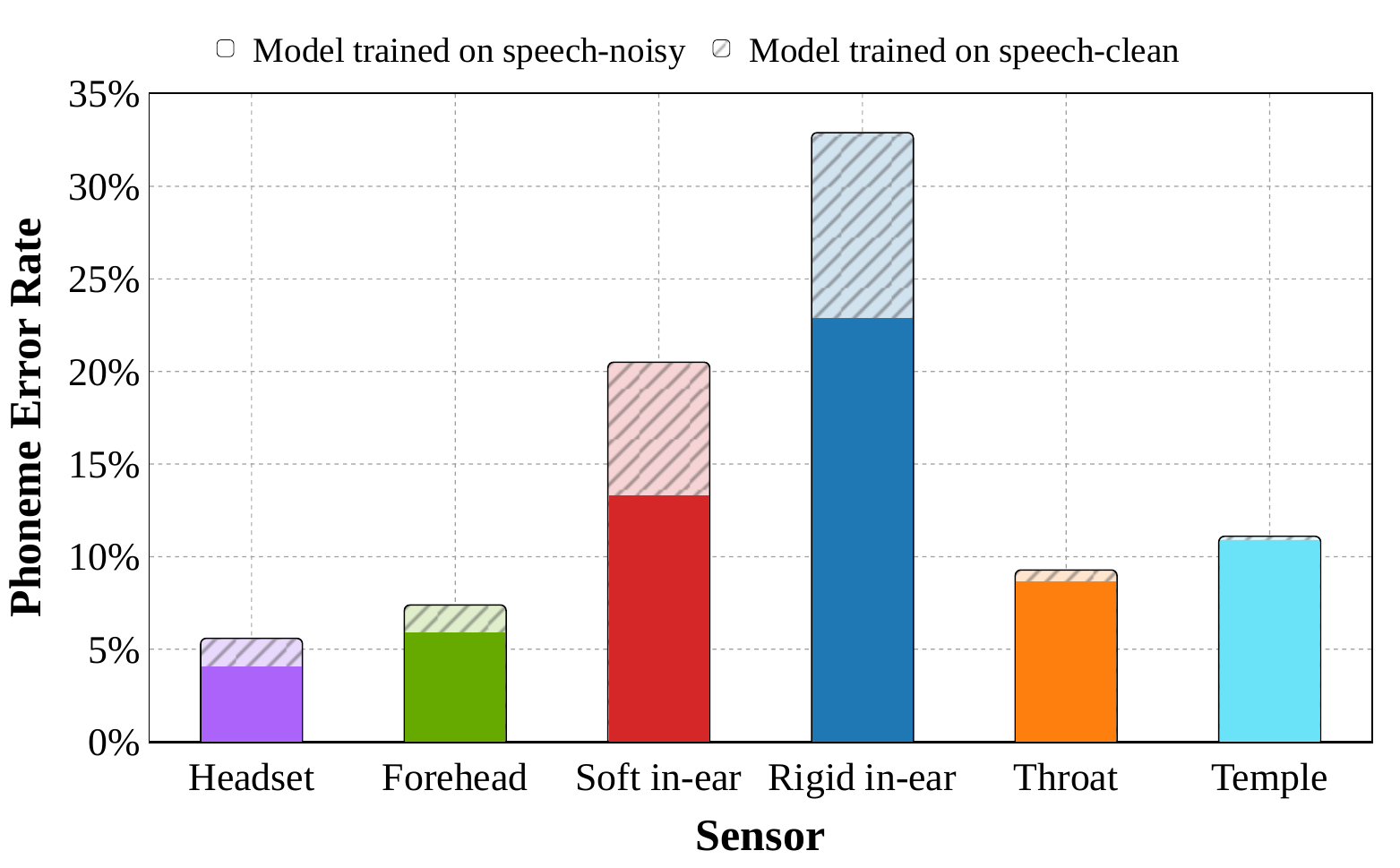}
  \caption{PER of the test recordings phonemized by the models trained on \texttt{speech-clean} and \texttt{speech-noisy} in noisy conditions}
  \label{img:per_clean_noisy}
\end{figure}

Figure~\ref{img:per_clean_noisy} presents the \texttt{speech-noisy} test results for phonemizers trained exclusively on the \texttt{speech-clean} subset, along with those refined using \texttt{speech-noisy} training data. These results highlight the impact of signal-to-external-noise ratios on the final performance of the phonemizers, as well as the performance gap between \texttt{speech-clean} and \texttt{speech-noisy} models. Specifically, as microphone-captured noise increases, the task becomes more challenging, and the domain gap between quiet and noisy conditions broadens. However, fine-tuning helps reduce this gap. These findings offer a pseudo-ranking of speech intelligibility in a 90 dB noisy environment across different microphones. Notably, the headset microphone remains the top performer due to its strong directivity toward the speaker's mouth. Since no processing was applied to reconstruct the bandwidth of body-conducted signals, high-frequency phonemes were lost. A significant performance drop is observed for the in-ear microphone in noisy conditions, likely due to inadequate acoustic sealing for certain ear canal anatomies that do not fit the sensor properly. While a "one-size-fits-all" design is ideal in theory, molded eartips would provide a better fit in practice, although this would require considerable additional manufacturing time for each participant. Throat and temple \texttt{speech-clean} phonemizers remain performant in noise, reflecting the high signal-to-external noise ratio in Table \ref{tab:specifications} of those sensors.

\vspace{-5mm}

\subsection{Speaker Verification}

This final section addresses speaker verification, which ascertains whether two audio files feature the same speaker, providing a boolean answer. Speaker verification constitutes a subset of speaker recognition tasks \cite{reynolds2008text}, which also encompass speaker identification and diarization. This task is especially pertinent in radio communication scenarios with multiple speakers, where BCMs can exacerbate the probability of voice confusion. Such confusion can impede fluid conversations, particularly in noisy environments. Furthermore, robust speaker identification systems are essential in contexts where heavy machinery is operated via voice commands, ensuring rigorous safety protocols are upheld. As stated by \cite{schwartz2018effects}, \textit{"The presence and use of high-frequency information from the speech signal improves the recognition of individual speakers, particularly in noisy environments."}. Therefore, as highlighted by previous research research\cite{sahidullah2016robust,sahidullah2017robust, liu2018vocal, shang2019enabling, gao2021voice}, the lack of these frequencies presents a significant challenge to BCMs speaker verification, necessitating further investigation to produce a more robust technology for these sensors. We therefore assess the efficiency of speaker verification tasks using body-conducted sensors from the Vibravox dataset. \\

\subsubsection{{Model}} The task is based on a distance metric comparison of speaker embedding results generated by a neural network from two distinct speech signals. Recently, Brydinskyi et al. \cite{brydinskyi2024comparison} investigated the effectiveness of various models trained on English datasets for differentiating non-English speakers. Their findings indicated superior performance from TitaNet \cite{koluguri2022titanet} and ECAPA-TDNN \cite{desplanques2020ecapa} compared to WavLM \cite{chen2022wavlm} or Pyannote \cite{bredin2023pyannote}. In light of the recent advances in speaker verification, particularly the ECAPA2 model \cite{thienpondt2023ecapa2}, which achieves state-of-the-art results, we elected to utilize its pre-trained version\footnote{\vspace{-5mm} available at \url{https://huggingface.co/Jenthe/ECAPA2}} on the Vibravox dataset for our analysis. \\

\subsubsection{{Testing procedure}}
It is crucial to highlight that the speaker verification task is only composed of a testing procedure, where we utilize a pretrained ECAPA2 model without any further training. Consequently, we solely utilize the test splits. The speaker verification testing set is constructed by forming pairs of speech signals, designated as $A$ and $B$, in alignment with the methodology outlined in \cite{brydinskyi2024comparison}. Notably, $A$ and $B$ may correspond to different airborne or body-conduction microphones. For each speaker, we generate $N_p$ pairs with their own recordings and an additional randomly selected $N_p$ pairs with recordings from different speakers. To increase the difficulty of the task, we maintain the possibility of separating speakers by gender. Here, $N_p$ represents the number of possible pairs for the speaker with the fewest recordings, which is set at 2346. We thus have a total of 98532 test pairs, given the fact that the \texttt{speech-clean} test set is composed of 21 speakers. We generate pairs only for the \texttt{speech-clean} subset to evaluate the model under quiet conditions and reuse these same pairs for the noisy condition. The latter is handled through a static mixing of \texttt{speech-clean} with \texttt{speechless-noisy}, which was shown in Section~\ref{subsec:se} to be a suitable modeling approach. This allows for a direct comparison of results between the two recording conditions. When comparing pairs of signals, we compute the cosine similarity and the Euclidean distance between their normalized embeddings. After inference on the entire set of pairs, we compute the Equal Error Rate (EER) and the normalized minimum of the Detection Cost Function (DCF) \cite{nist2018speaker} from the cosine similarity scores. \\

\subsubsection{{Experiments in quiet conditions}}

\begin{figure}[ht!]
  \centering
  \includegraphics[width=0.98 \linewidth]{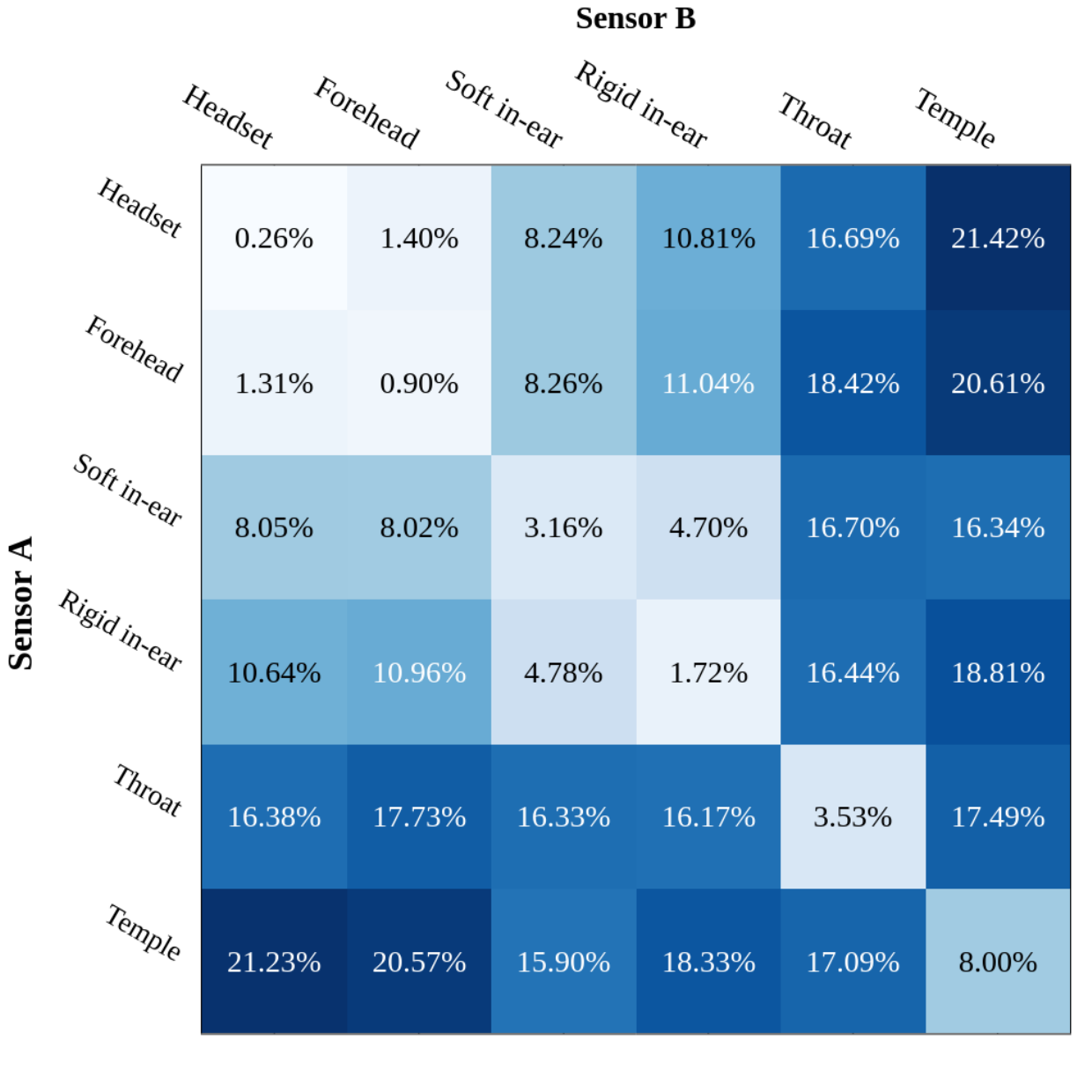}
  \caption{EER on \texttt{speech-clean} with the speaker verification model for all pairs}
  \label{img:eer_matrix_speech_clean}
\end{figure}

The results of the ECAPA2 model test are presented in Figure \ref{img:eer_matrix_speech_clean} for all sensor combinations. As with speech-to-phonemes, the matrix exhibits a diagonal with lower coefficients because it is easier to distinguish between two recordings taken from the same audio sensor. The asymmetry of the matrix is to be expected, given that the script for generating pairs is not commutative due to its randomness. Nevertheless, the results from the upper triangular and lower triangular parts are very close and show a low variance. It is immediately apparent that there is a considerable discrepancy in the diagonal EER between different sensors. To illustrate this point, the headset and temple vibration pickup exhibit an EER differing by a rate of 30. The cross-tested EER for the forehead/headset and soft/rigid in-ear microphones remains low compared to their initial performance, indicating the proximity of those sensors. Once more, we observe that the difficulty of the task is again mostly driven by the sensor's bandwidth. \\

\vspace{-1mm}

\subsubsection{{Effect of EBEN enhancement in quiet conditions}}
\begin{figure}[h!]
  \centering
  \includegraphics[width=0.98 \linewidth]{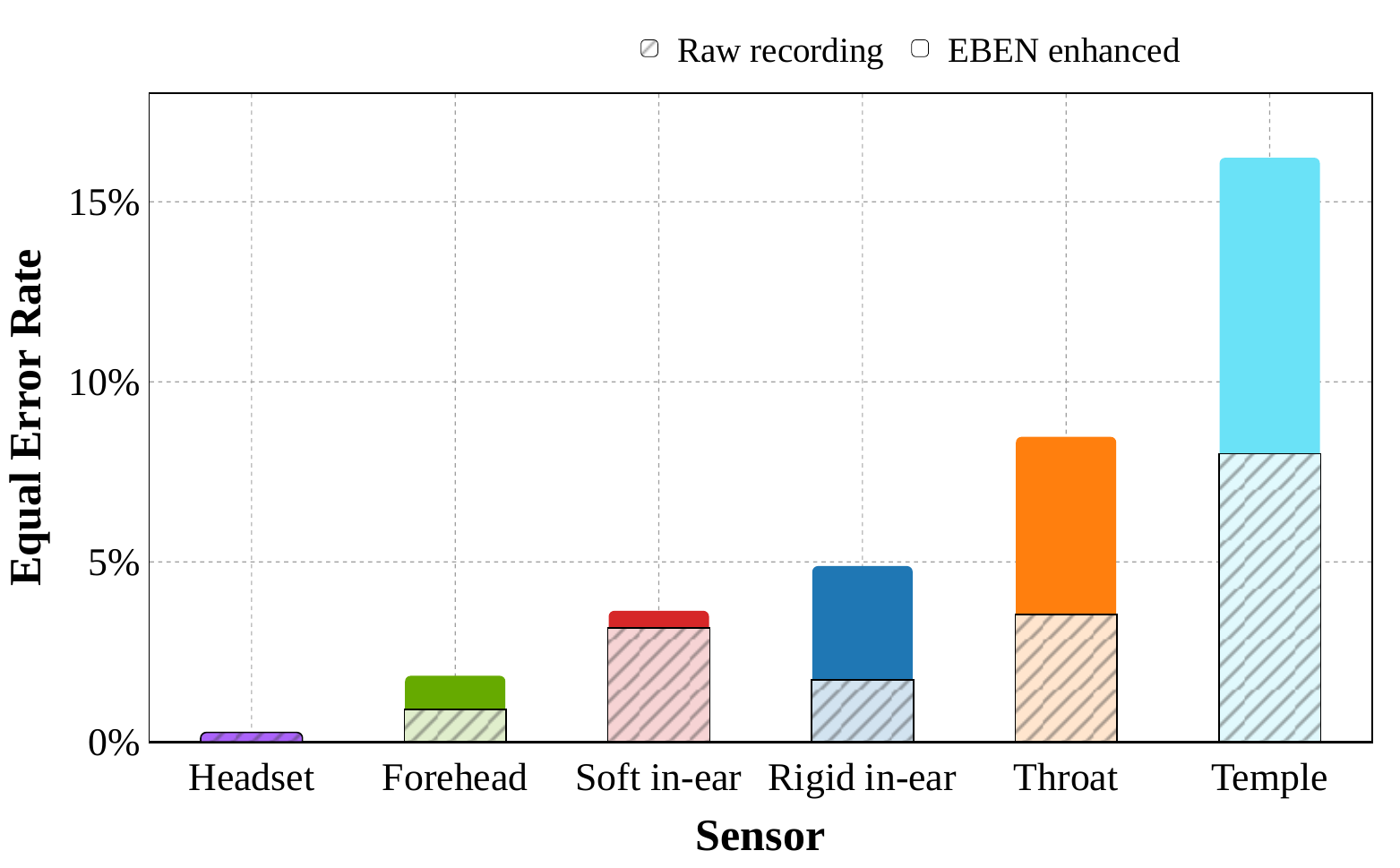}
  \caption{EER on \texttt{speech-clean} of the raw and EBEN-enhanced recordings with the speaker verification model on pairs of the same audio sensor}
  \label{img:eer_comparison_histogram}
\end{figure}

To complete this study, the Equal Error Rate (EER) was compared between raw and EBEN-enhanced signals. Contrary to the observations in Section \ref{subsec:stt}, a deterioration in performance is evidenced in Figure \ref{img:eer_comparison_histogram}. Note that speaker identity preservation was not included among the three losses—adversarial loss ($\mathcal{L}_\mathcal{G}^{adv}$), feature loss ($\mathcal{L}_\mathcal{G}^{feat}$), and spectral loss ($\mathcal{L}_\mathcal{G}^{spec}$) which were employed to adjust the EBEN parameters. One possible explanation for the observed performance drop is that the ECAPA2 model may interpret the EBEN outputs as being out-of-distribution, which could account for the poor performance. This issue could be addressed in future work by incorporating speaker identity preservation losses into the overall objective, as explored in \cite{wang2022speaker}.

\subsubsection{{Experiments in noisy conditions}}
\begin{figure}[h!]
  \centering
  \includegraphics[width=0.98 \linewidth]{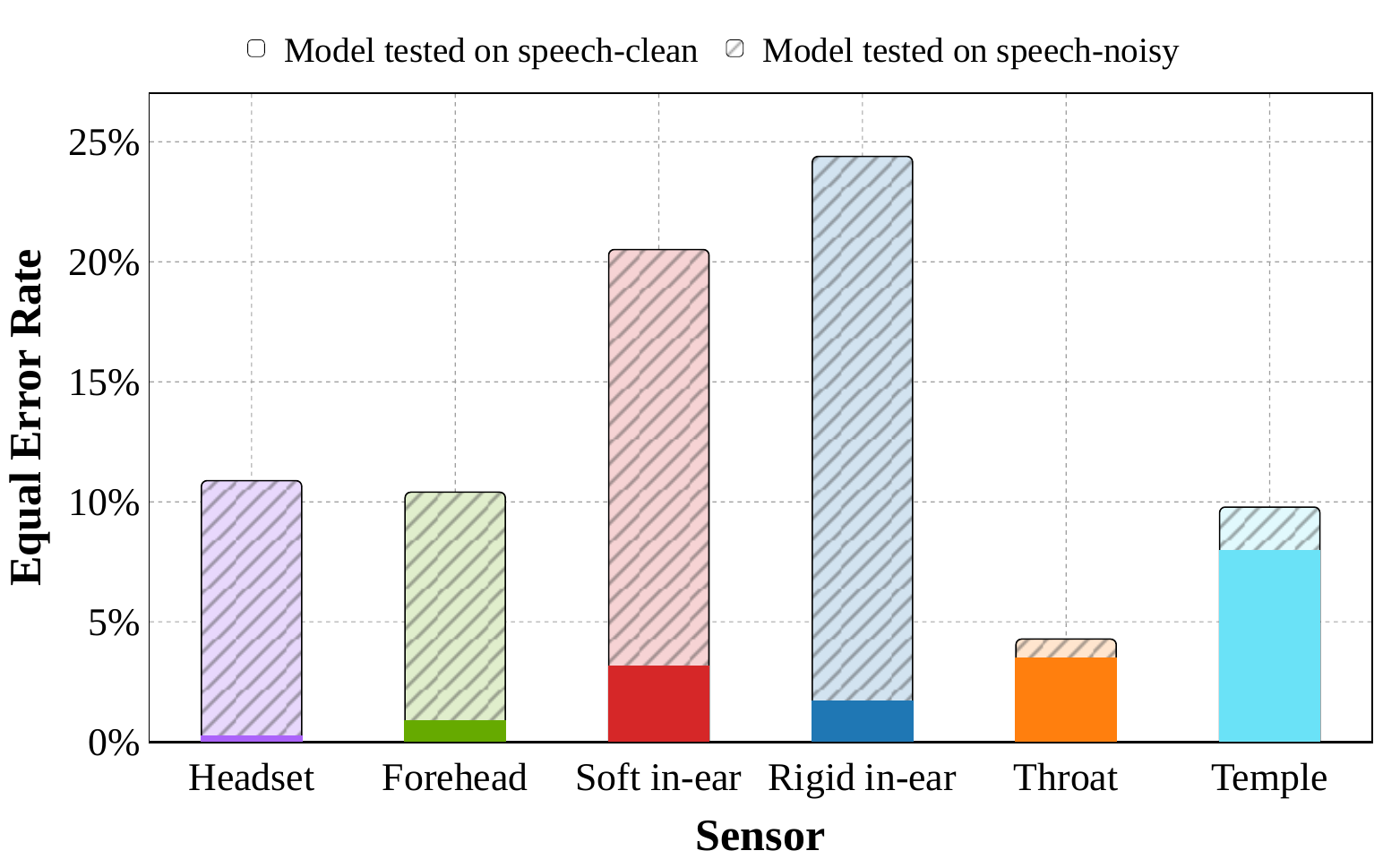}
  \caption{EER on \texttt{speech-clean} and synthetic \texttt{speech-noisy} with the speaker verification model on pairs of the same audio sensor}
  \label{img:eer_noisy}
\end{figure}

This section evaluates ECAPA2's performance on both the \texttt{speech-clean} and the synthetic mix \texttt{speech-clean} + \texttt{speechless-noisy} datasets, the latter simulating \texttt{speech-noisy} while preserving identical pairs. Figure~\ref{img:eer_noisy} shows that the Equal Error Rate (EER) generally increases, although the extent of degradation varies. For the two microphones most resistant to noise—the throat and temple microphones—the performance decline is less significant. In contrast, the in-ear microphones, affected by acoustic sealing issues, exhibit a more substantial drop. It is noteworthy that raw temple and throat signals outperform those captured by the noise-resilient airborne microphone -- even without bandwidth extension or denoising -- demonstrating their potential for communication in such conditions. The raw throat signal outperforms the headset microphone signal exclusively in the speaker verification context, likely due to the low-pass filtering properties related to speaker anatomy, which may facilitate speaker identification, while the high-frequency components are more crucial for accurate phoneme recognition.

\section{Discussion and Future Work}
\label{sec:conclusion}
Building the Vibravox dataset has been a protracted undertaking that has necessitated the achievement of several milestones. These include the construction of equipment, the development of necessary software, the recording of participants, the filtering of data, and other related tasks. This highly curated dataset, comprising four subsets —\texttt{speech-clean}, \texttt{speech-noisy}, \texttt{speechless-clean}, and \texttt{speechless-noisy} — is now publicly available. The total duration of the 6-channel recordings is over 45 hours per sensor, with 188 participants. This corpus is being made available online on HuggingFace via Creative Commons Attribution 4.0 International license. Using Vibravox has proven to yield satisfactory results in Speech Enhancement, Speech-to-Phonemes and Speaker Verification, which are three fundamental speech processing tasks. The EBEN models used for speech enhancement demonstrated the capacity to enhance objective metrics such as Noresqa-MOS and STOI while simultaneously improving phoneme transcription on every sensor. However, it was found to have a detrimental impact on speaker verification performance. The aforementioned tasks further contributed to a deeper understanding of the distinctive acoustic subtleties of the various audio sensors and their respective placements.

We hope this release will drive progress in body-conducted speech analysis and support the development of robust communication systems for real-world applications.

\section*{Acknowledgements}

The authors would like to express their gratitude to the 200 participants in the Vibravox measurement protocol, whose invaluable contributions made this research possible. We would also like to thank Jean-Baptiste Doc for his assistance in developing 3D-printed components that facilitated sensor positioning for participants with diverse morphologies and Philippe Chenevez for his expertise in electrical engineering for the design of pre-amplifier circuits. This work was granted access to the HPC/AI resources of [CINES / IDRIS / TGCC] under the grant 2022-AD011013469 awarded by GENCI and partially funded by the French National Research Agency under the ANR Grant No. ANR-20-THIA-0002.

\section*{Ethical Considerations}

In accordance with the guidelines of the General Data Protection Regulation (GDPR), which governs the processing and protection of personal data of citizens and residents of the European Union (EU), all necessary measures were taken to protect the privacy and rights of participants in this study. This included obtaining informed consent, clearly communicating the purpose of data collection, and ensuring anonymization and encryption of data during storage and analysis. In addition, the use of Wikipedia text ensures compliance with copyright and licensing issues. Finally, to increase the representativeness and inclusivity of the dataset, a deliberate effort was made to recruit a diverse and gender-balanced group of participants.

\section*{Declaration of AI-assisted technologies in the writing process}

In the preparation of this work, the authors utilized QuillBot's Grammar Checker and Mistral's LeChat to improve the quality of the English language in the manuscript. Following the use of these tools, the authors thoroughly reviewed and edited the content as necessary and assume full responsibility for the final version of the publication.

\ifCLASSOPTIONcaptionsoff
  \newpage
\fi



%

\bibliographystyle{IEEEtran}
\bibliography{vibravox.bib}

\clearpage

\begin{IEEEbiography}[{\includegraphics[width=1in,height=1.25in,clip,keepaspectratio]{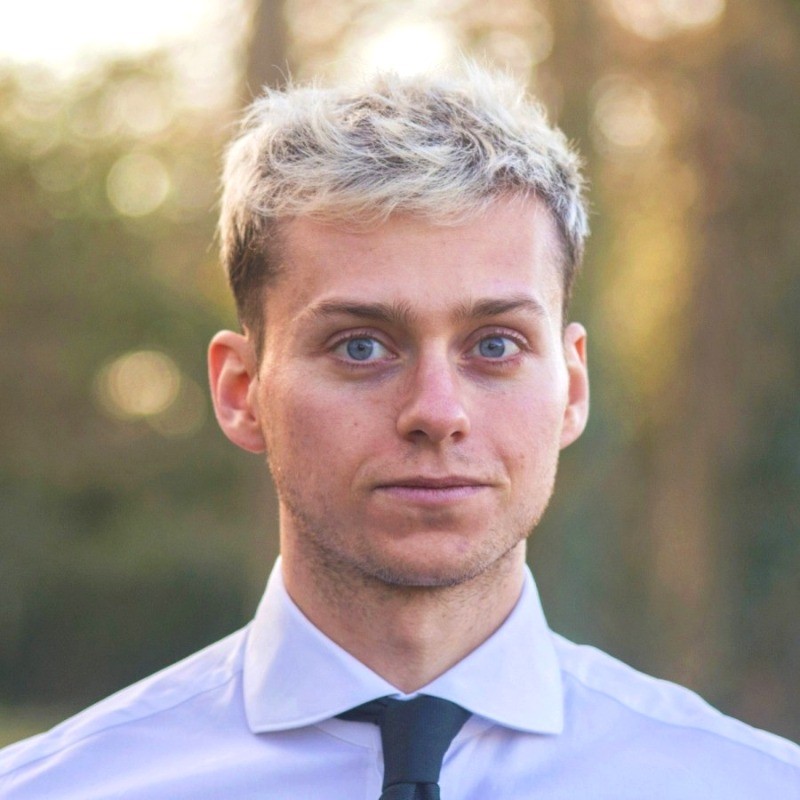}}]{Julien Hauret}
is a PhD candidate at Cnam Paris, pursuing research in machine learning applied to speech processing. He holds two MSc degrees from ENS Paris Saclay, one in Electrical Engineering (2020) and the other in Applied Mathematics (2021). His research training is evidenced by his experiences at Columbia University, the French Ministry of the Armed Forces and the Pulse Audition start-up. Additionally, he has lectured for two consecutive years on algorithms and data structures at the École des Ponts ParisTech. His research focuses on the use of deep learning for speech enhancement applied to body-conducted speech. With a passion for interdisciplinary collaboration, Julien aims to improve human communication through technology.
\end{IEEEbiography}

\begin{IEEEbiography}[{\includegraphics[width=1in,height=1.25in,clip,keepaspectratio]{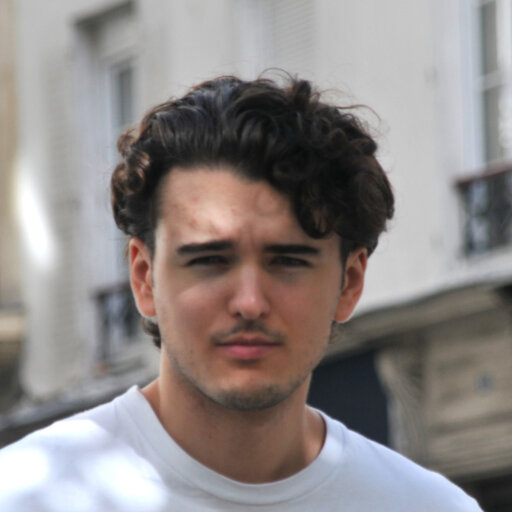}}]{Malo Olivier}
completed an internship in the LMSSC lab at Cnam Paris during his studies at INSA Lyon. He graduated from the Computer Science Department at INSA Lyon in June 2024 and he now works at Cnam as a research engineer on an experimental project. With skills in implementing solutions—from information systems challenges to deep neural network architectures, including web applications—he foresees pursuing a PhD in Artificial Intelligence, specializing in deep neural networks applied to scientific fields, hoping his engineering profile highlights his implementation abilities in high-impact projects.
\end{IEEEbiography}

\begin{IEEEbiography}[{\includegraphics[width=1in,height=1.25in,clip,keepaspectratio]{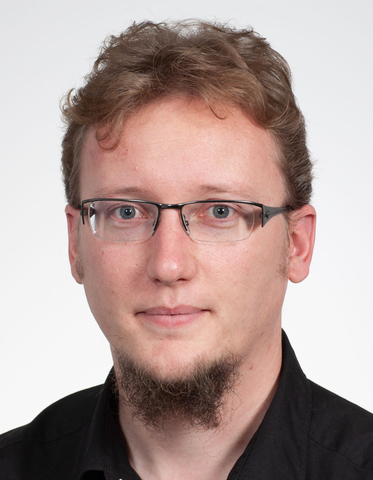}}]{Thomas Joubaud}
is a Research Associate at the Acoustics and Soldier Protection department within the French-German Research Institute of Saint-Louis (ISL), France, since 2019. In 2013, he received the graduate degree from Ecole Centrale Marseille, France, as well as the master’s degree in Mechanics, Physics and Engineering, specialized in Acoustical Research, of the Aix-Marseille University, France. He earned the Ph.D. degree in Mechanics, specialized in Acoustics, of the Conservatoire National des Arts et Métiers (Cnam), Paris, France, in 2017. The thesis was carried out in collaboration with and within the ISL. From 2017 to 2019, he worked as a post-doctorate research engineer with Orange SA company in Cesson-Sévigné, France. His research interests include audio signal processing, hearing protection, psychoacoustics, especially speech intelligibility and sound localization, and high-level continuous and impulse noise measurement.
\end{IEEEbiography}

\begin{IEEEbiography}[{\includegraphics[width=1in,height=1.25in,clip,keepaspectratio]{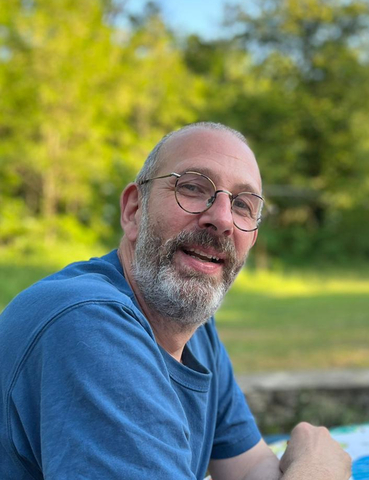}}]{Christophe Langrenne}
is a Research Engineer at the Laboratoire de Mécanique des Structures et des Systèmes Couplés (LMSSC) at the Conservatoire National des Arts et Métiers (Cnam), Paris, France. After completing his PhD on the regularization of inverse problems, he developed a fast multipole method (FMM) algorithm for solving large-scale scattering and propagation problems. Also interested in 3D audio, he co-supervised 3 PhD students on this theme, in particular on Ambisonic (recording and decoding) and binaural restitution (front/back confusions).
\end{IEEEbiography}

\begin{IEEEbiography}[{\includegraphics[width=1in,height=1.25in,clip,keepaspectratio]{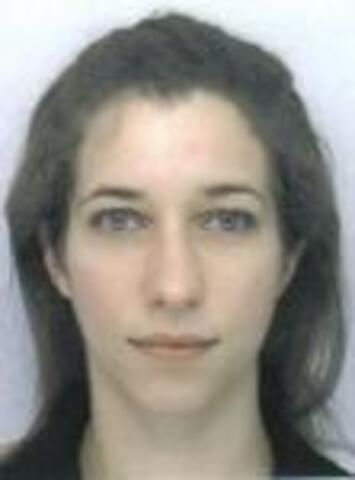}}]{Sarah Poiree}
is a technician at the Laboratoire de Mécanique et des Systèmes Couplés (LMSSC) within the Conservatoire National des Arts et Métiers (Cnam), Paris, France. Her activities focus on the design and development of experimental setups. Notably, she contributed to the creation of the 3D sound spatialization system used during the recording of the Vibravox dataset.
\end{IEEEbiography}

\begin{IEEEbiography}[{\includegraphics[width=1in,height=1.25in,clip,keepaspectratio]{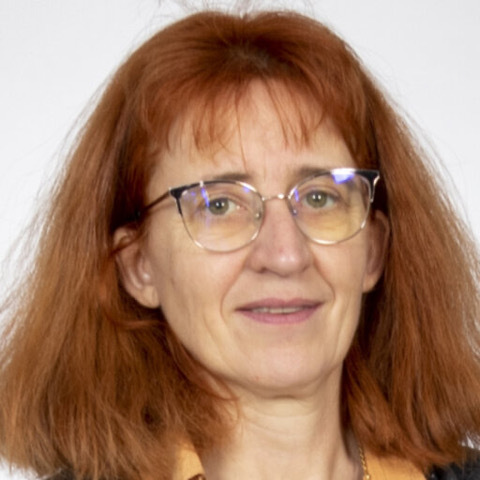}}]{Véronique Zimpfer}
is a Scientific Researcher at the Acoustics and Soldier Protection department within the French-German Research Institute of Saint-Louis (ISL), Saint-Louis, France, since 1997. She holds a M.Sc in Signal Processing from the Grenoble INP, France and obtained a PhD in Acoustics from INSA Lyon, France, in 2000. Her expertise lies at the intersection of communication in noisy environments and auditory protection. Her research focuses on improving adaptive auditory protectors, refining radio communication strategies through unconventional microphone methods, and enhancing auditory perception while utilizing protective gear.
\end{IEEEbiography}

\begin{IEEEbiography}[{\includegraphics[width=1in,height=1.25in,clip,keepaspectratio]{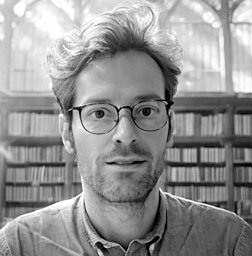}}]{Éric Bavu}
is a Full Professor of Acoustics and Signal Processing at the Laboratoire de Mécanique des Structures et des Systèmes Couplés (LMSSC) within the Conservatoire National des Arts et Métiers (Cnam), Paris, France. He completed his undergraduate studies at École Normale Supérieure de Cachan, France, from 2001 to 2005. In 2005, he earned an M.Sc in Acoustics, Signal Processing, and Computer Science Applied to Music from Université Pierre et Marie Curie Sorbonne University (UPMC), followed by a Ph.D. in Acoustics jointly awarded by Université de Sherbrooke, Canada, and UPMC, France, in 2008. He also conducted post-doctoral research on biological soft tissues imaging at the Langevin Institute at École Supérieure de Physique et Chimie ParisTech (ESPCI), France. Since 2009, he has supervised 8 Ph.D. students at LMSSC, focusing on time domain audio signal processing for inverse problems, 3D audio, and deep learning for audio. His current research interests encompass deep learning methods applied to inverse problems in acoustics, moving sound source localization and tracking, speech enhancement, and speech recognition.
\end{IEEEbiography}


%

\clearpage
\onecolumn

\appendices




\section{Dataset statistics}

\begin{table*}[h!]
\caption{Speakers age balance}
\centering
\begin{tabular}{@{}ccccc@{}}
\toprule
Gender & Mean age (years)  & Median age (years) &   Min age (years) & Max age (years) \\
\midrule
Female & 25.9  & 22   & 19    & 59  \\
Male & 31.4 &27& 18 & 82  \\
\midrule
\textbf{All} & \textbf{28.55} & \textbf{25}  & \textbf{18}   & \textbf{82} \\
\bottomrule
\end{tabular}
\end{table*}

\section{Speech Enhancement training curves on the \texttt{speech-clean} subset}

\begin{figure}[H]
  \centering
  \subfloat[Validation STOI]{\includegraphics[width=0.49\linewidth]{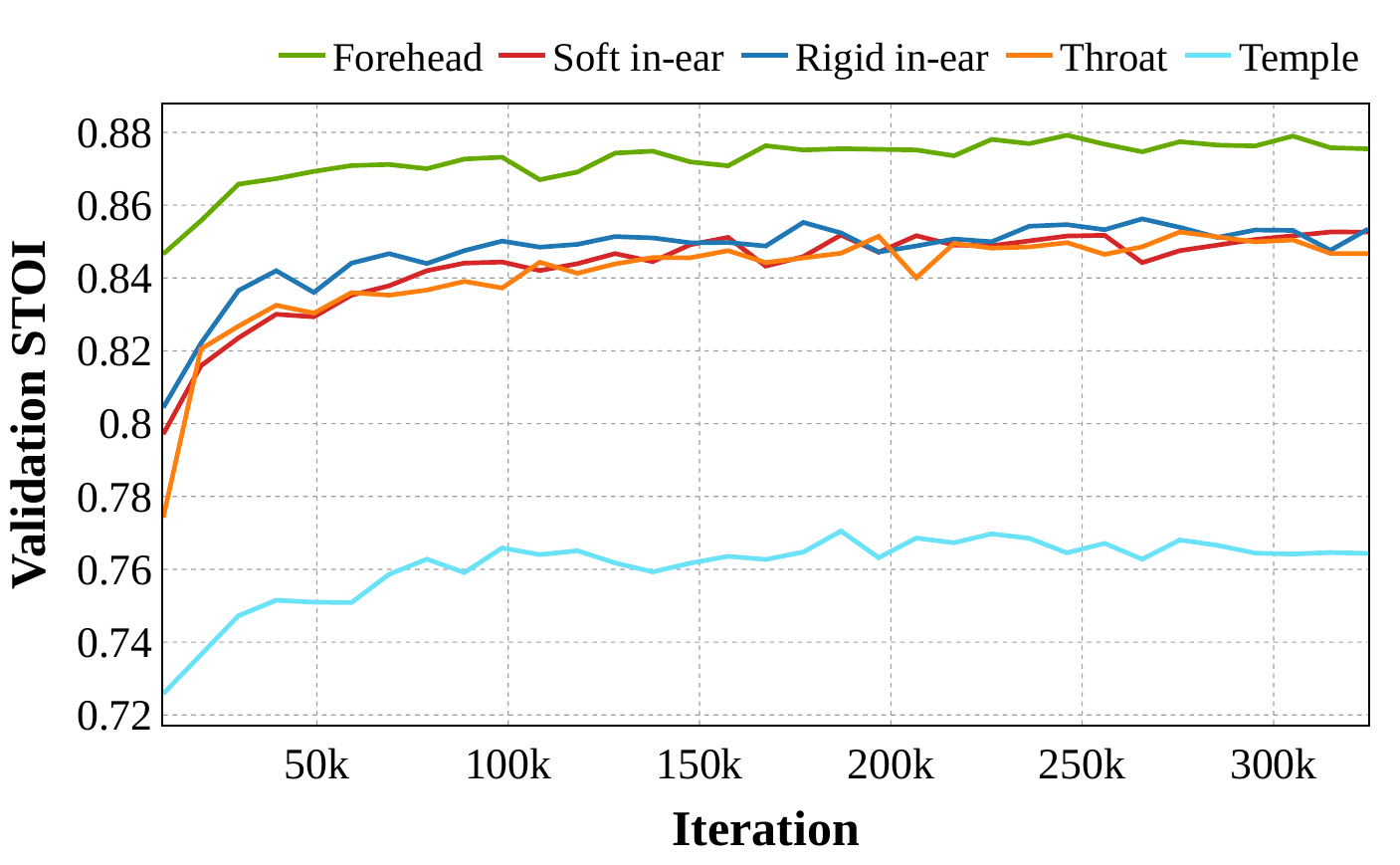}}
  \subfloat[Validation Noresqa-MOS]{\includegraphics[width=0.49\linewidth]{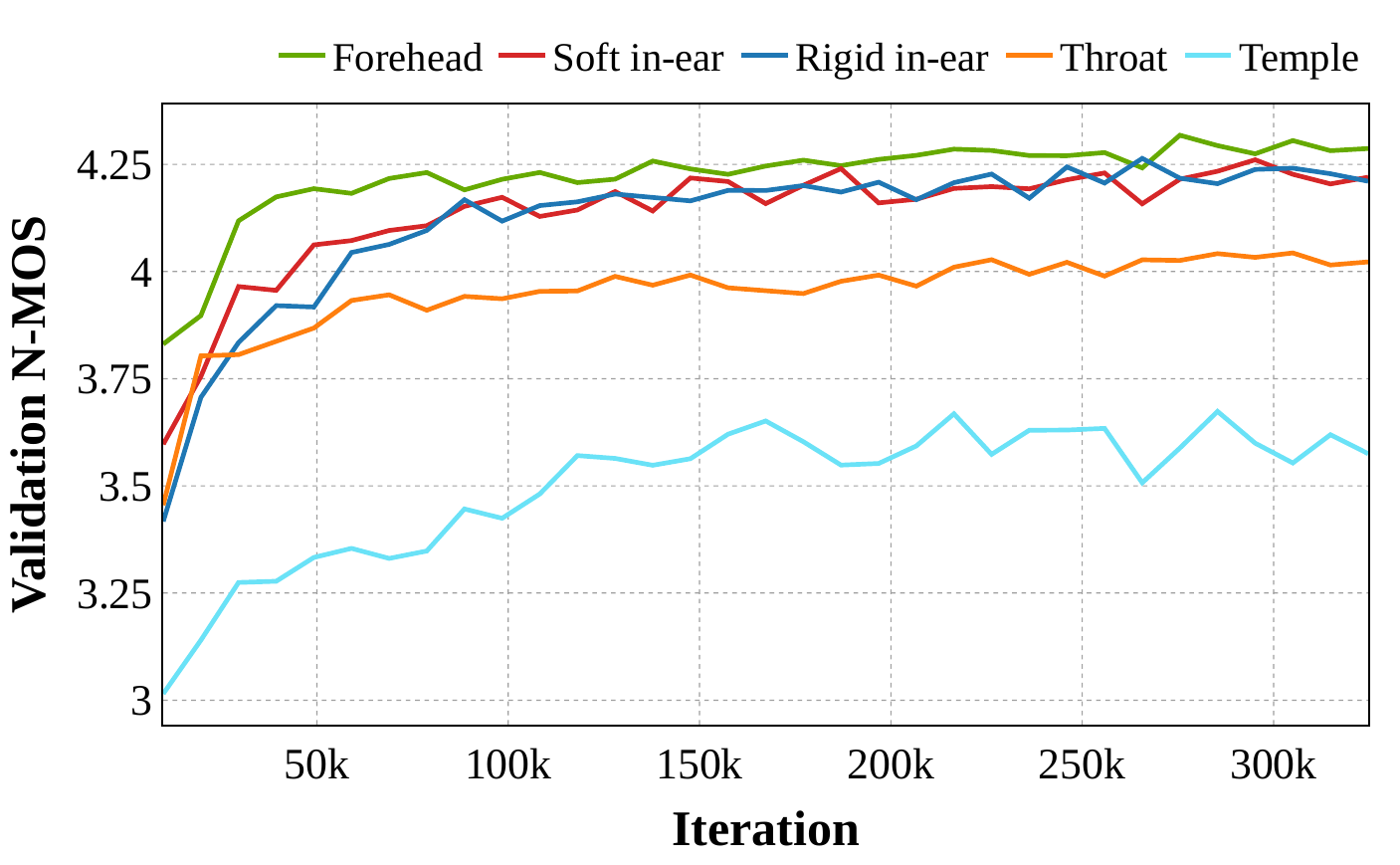}} \\
  \caption{EBEN validation curves during training on the \texttt{speech-clean} subset}
  \label{fig:appendix_bwe}
\end{figure}

\section{Speech-to-Phonemes training curves on the \texttt{speech-clean} subset}

\begin{figure}[H]
  \centering
  \subfloat[Validation Connectionist Temporal Classification Loss]{\includegraphics[width=0.49\linewidth]{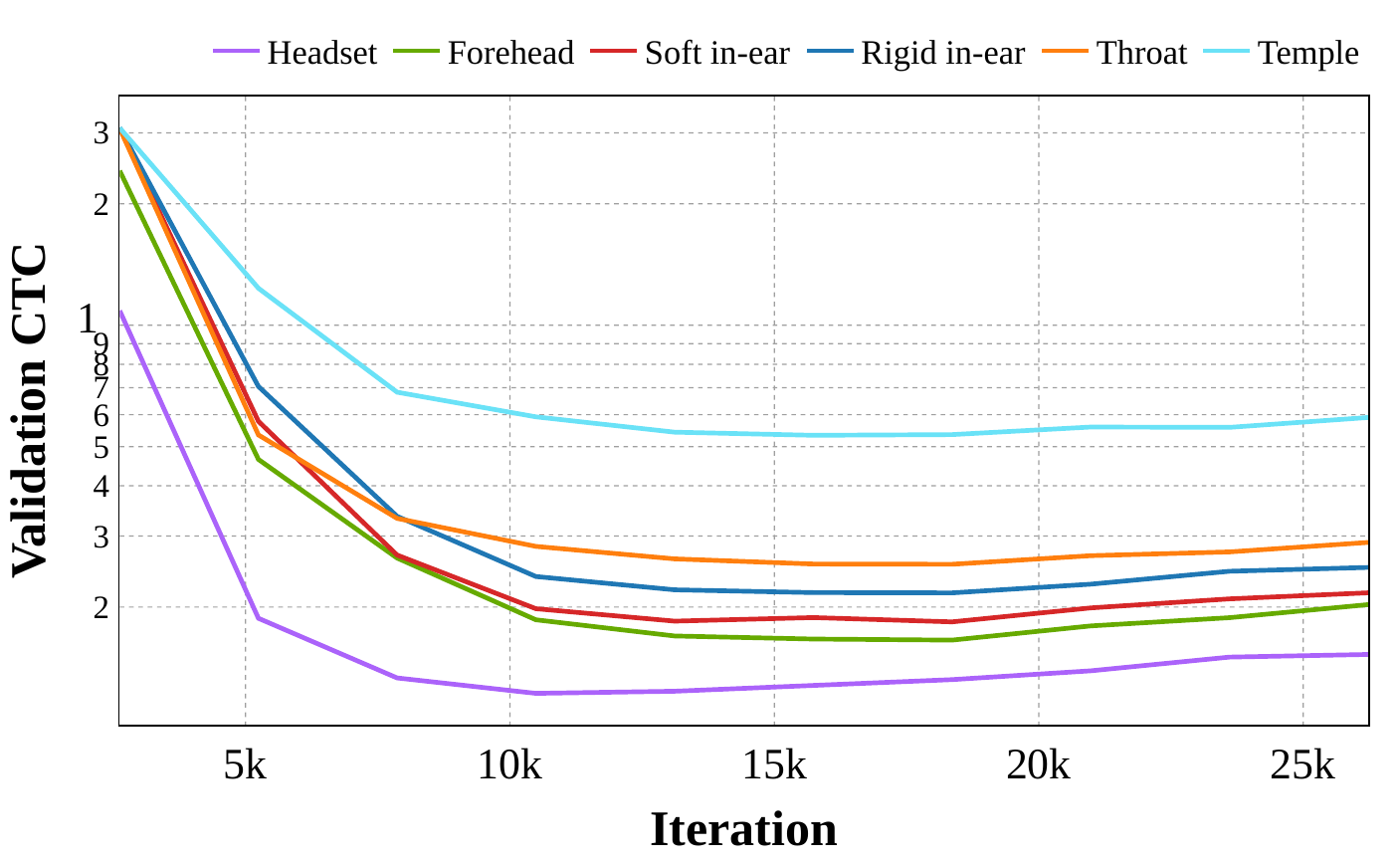}}
  \subfloat[Validation Phoneme-Error-Rate]{\includegraphics[width=0.49\linewidth]{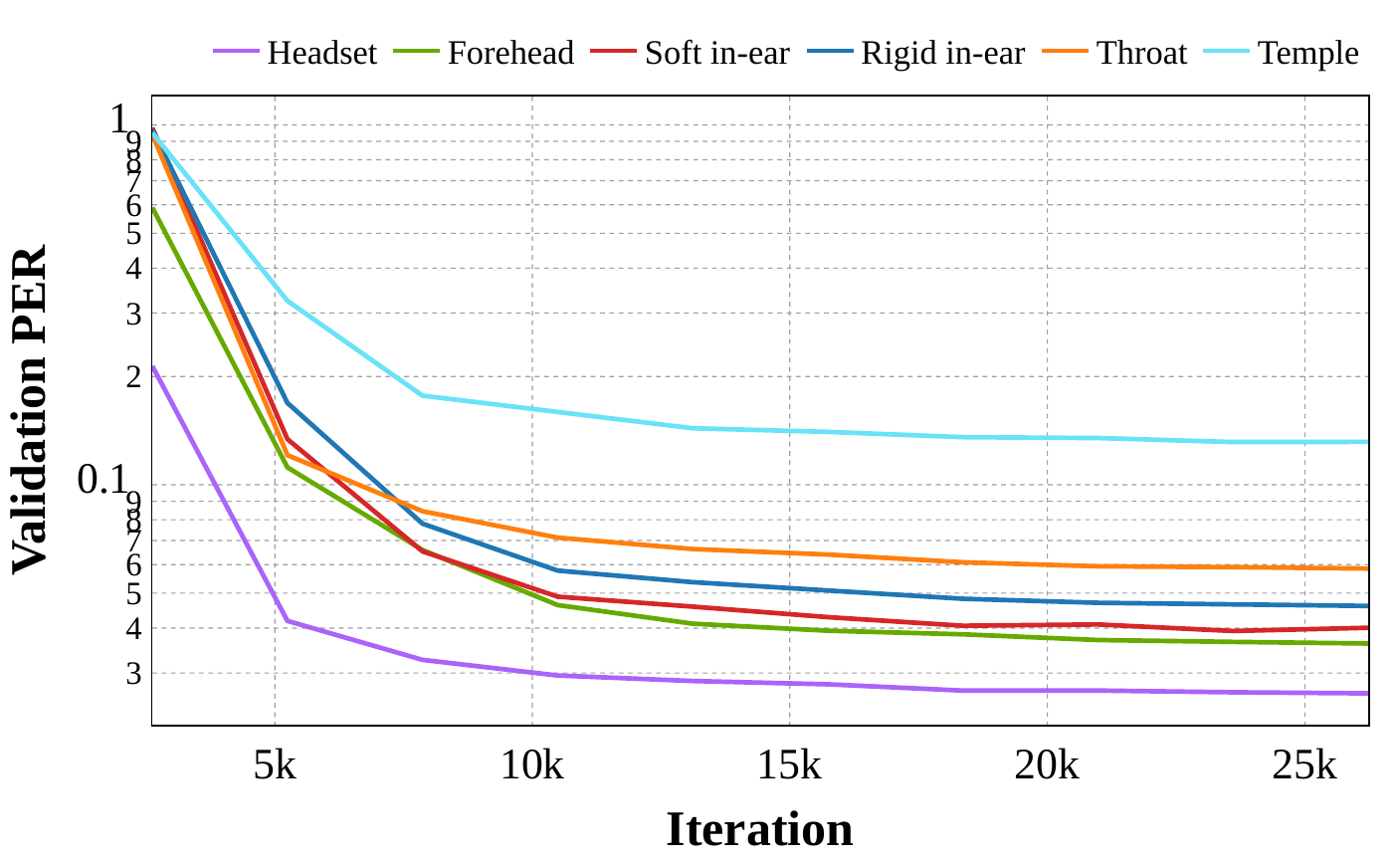}} \\
  \caption{Wav2vec2 validation curves during training on the \texttt{speech-clean} subset}
  \label{fig:appendix_stp}
\end{figure}

\newpage
\section{Coherence function statistics between body-conducted and airborne microphones on \texttt{speech-clean}}
\begin{figure*}[h]
\centering
    \subfloat[Forehead]{\includegraphics[width=0.49\linewidth]{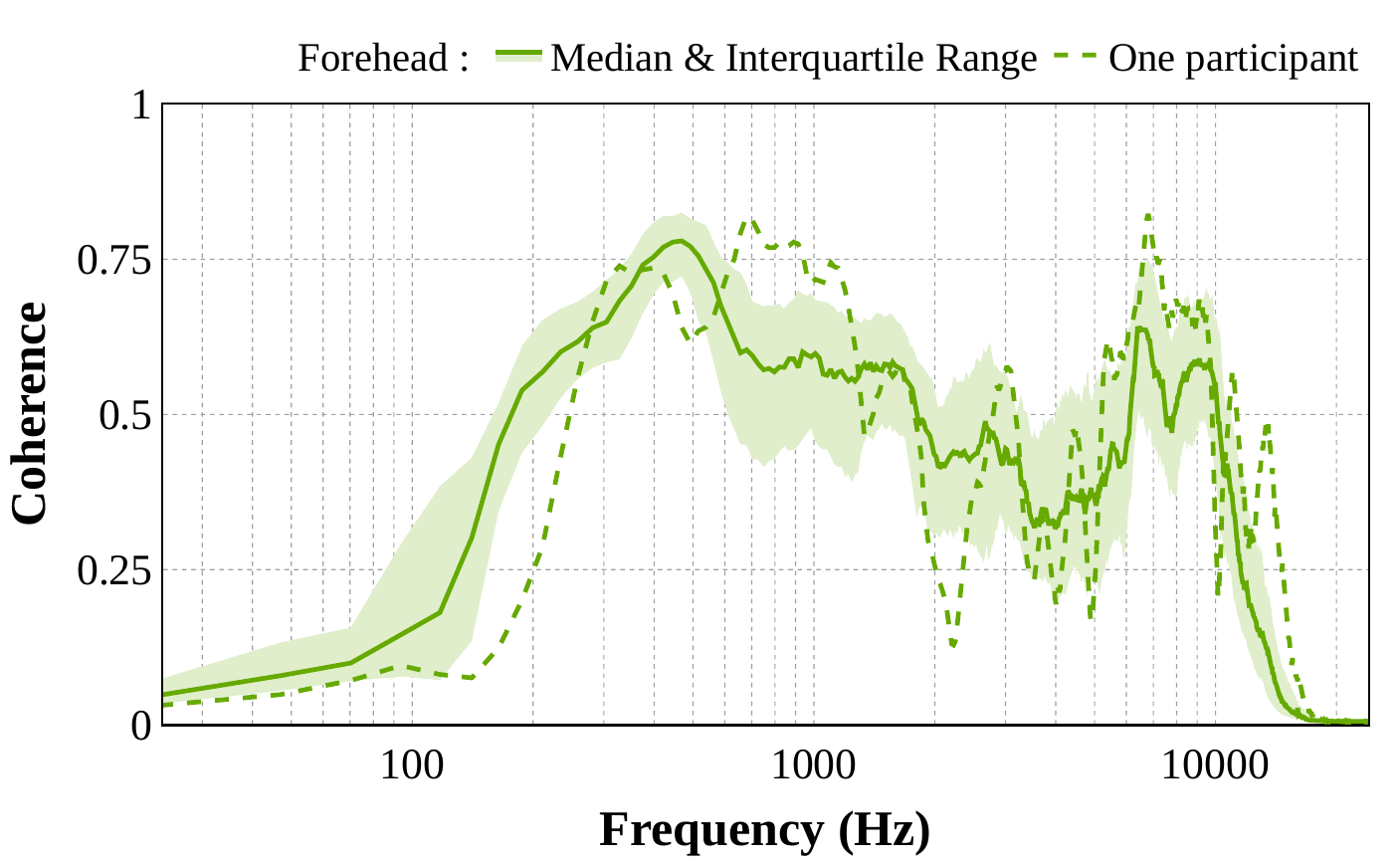}}
    \subfloat[Soft in-ear]{\includegraphics[width=0.49\linewidth]{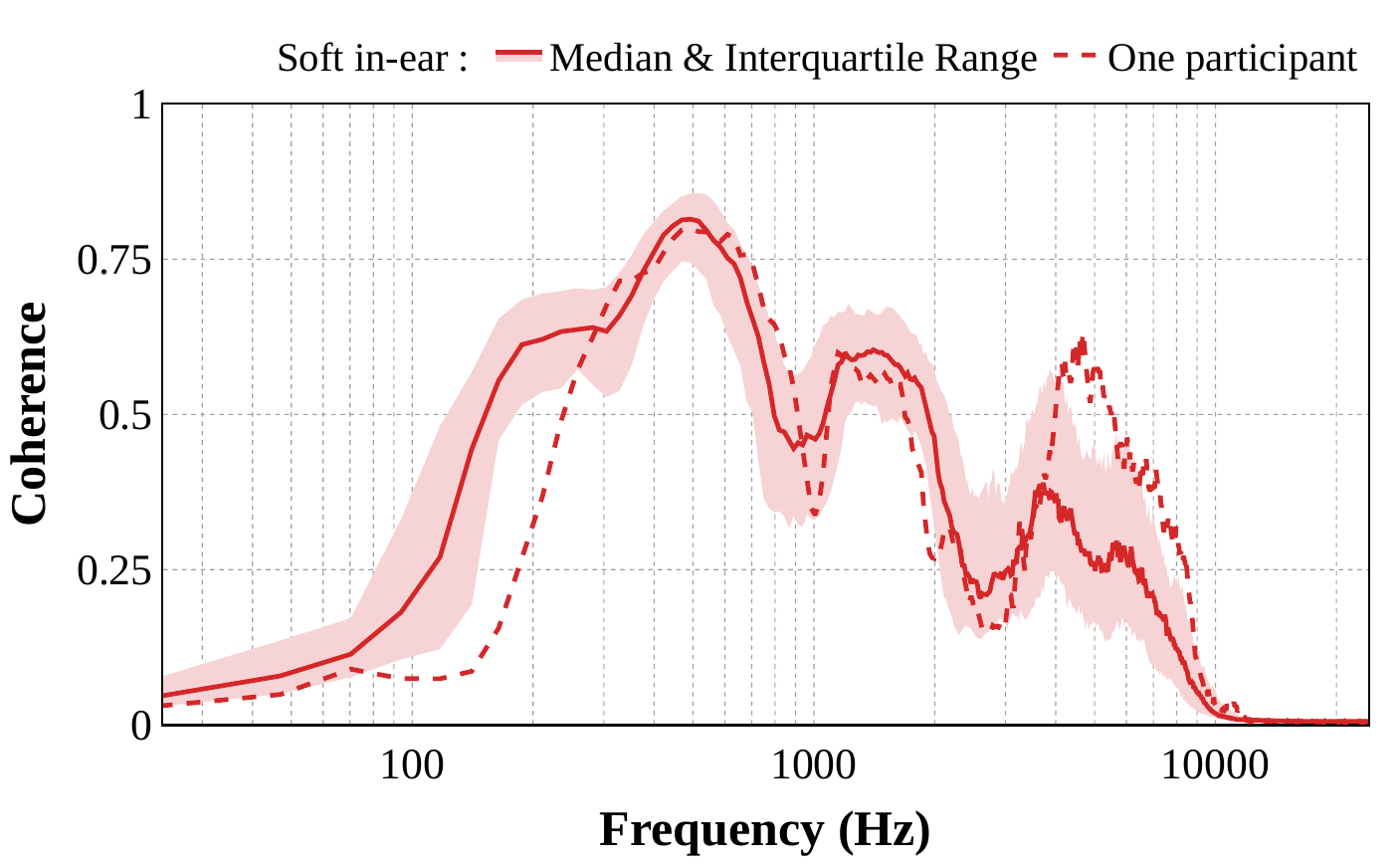}} \\
    \subfloat[Rigid in-ear]{\includegraphics[width=0.49\linewidth]{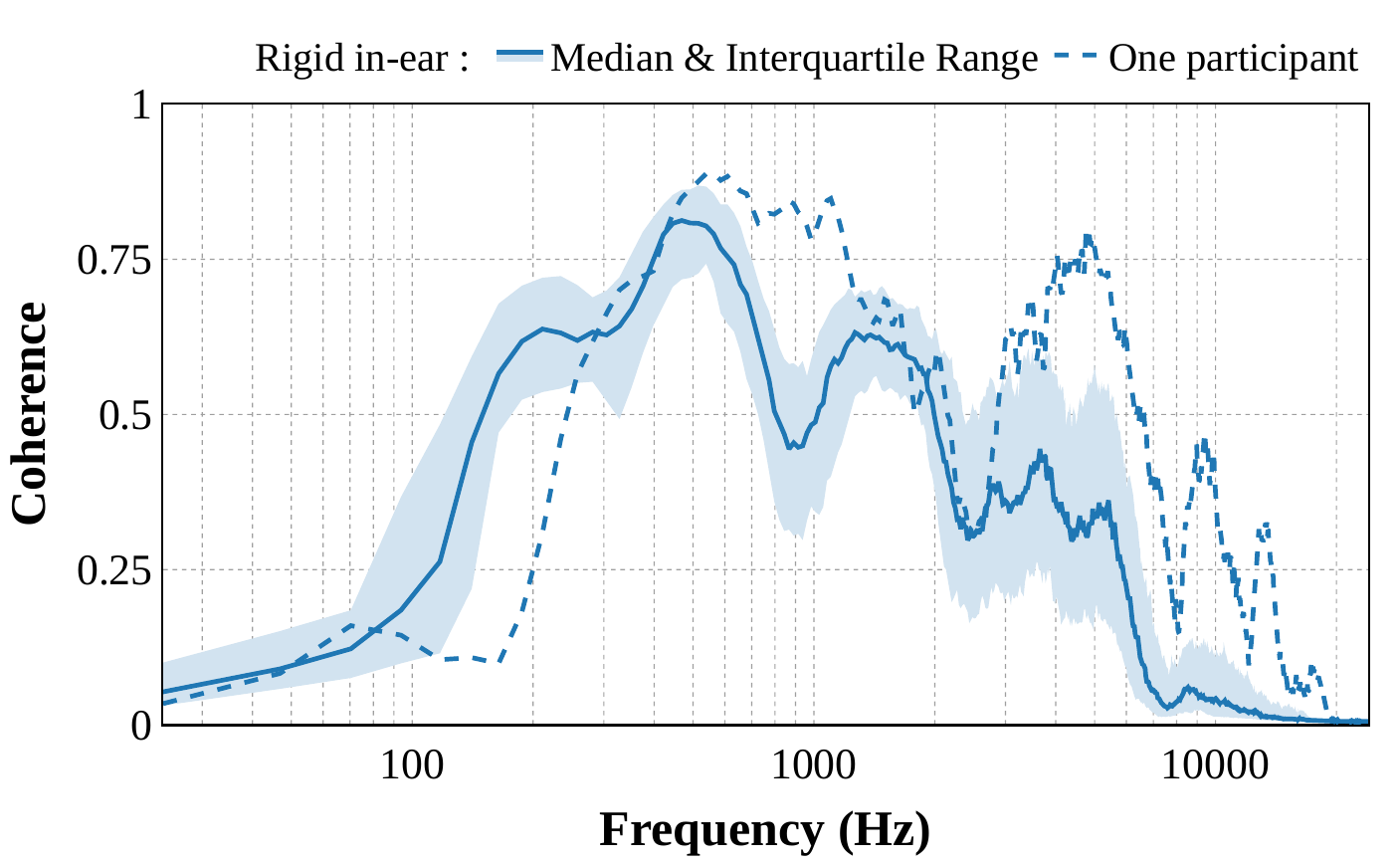}}
    \subfloat[Throat]{\includegraphics[width=0.49\linewidth]{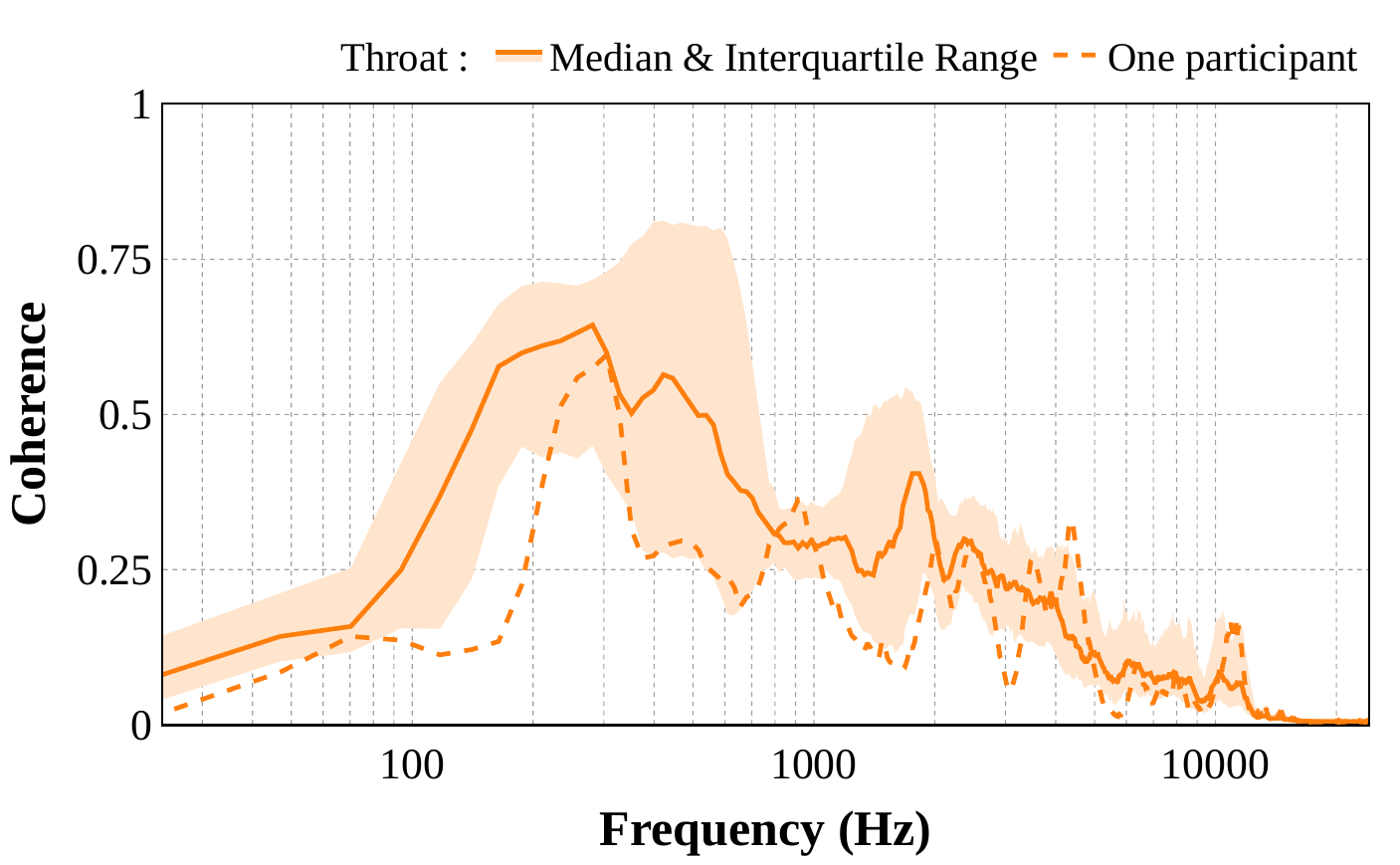}} \\
    \subfloat[Temple]{\includegraphics[width=0.49\linewidth]{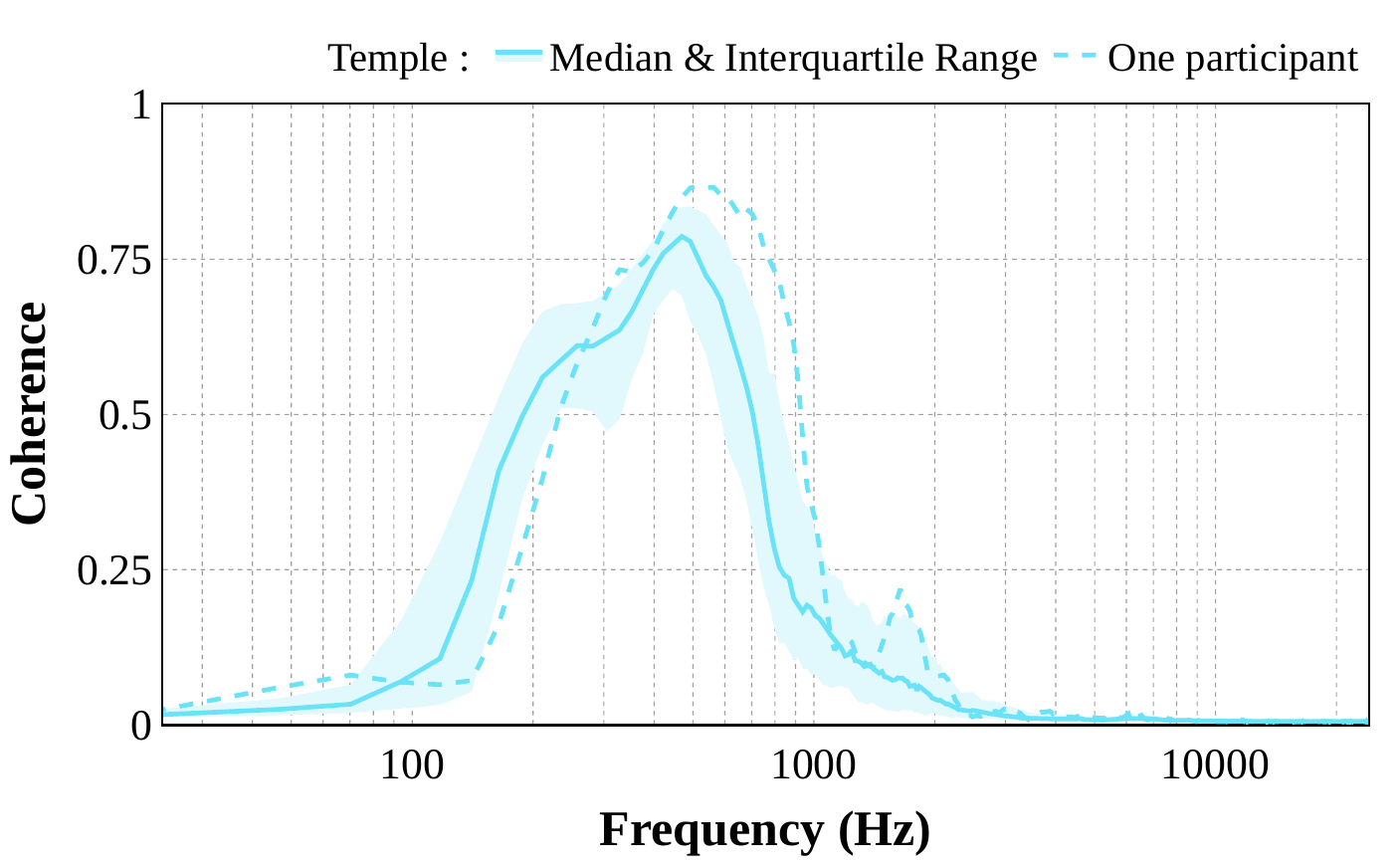}}
\caption{Median coherence function per sensor on \texttt{speech-clean} given with their 25-75$\%$ IQR and a single participant}
\label{fig:appendix_coherences}
\end{figure*}

\newpage
\section{Power spectral density of microphones on \texttt{speechless-clean}}
\begin{figure*}[h]
\centering
    \subfloat[Headset]{\includegraphics[width=0.49\linewidth]{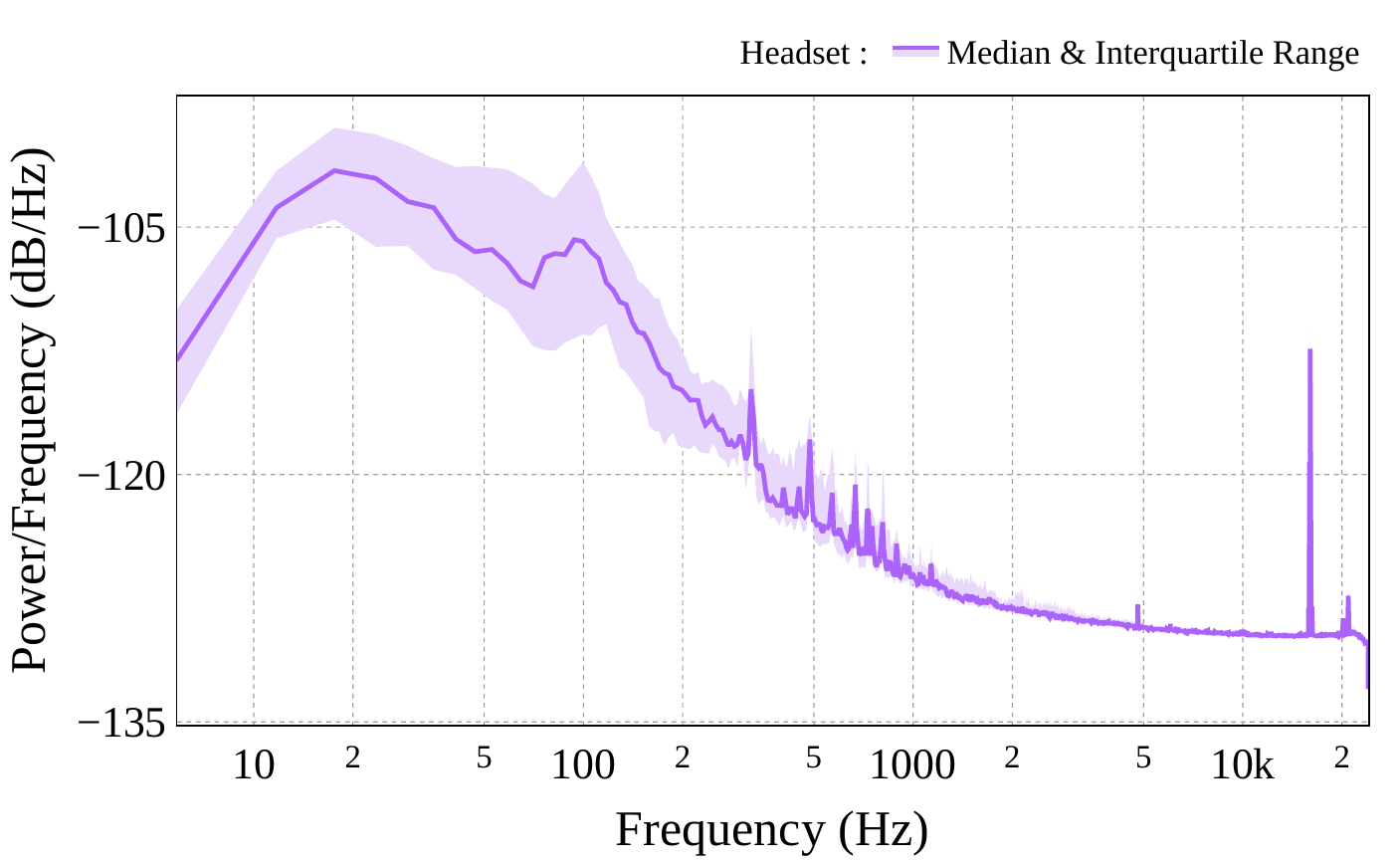}}
    \subfloat[Forehead]{\includegraphics[width=0.49\linewidth]{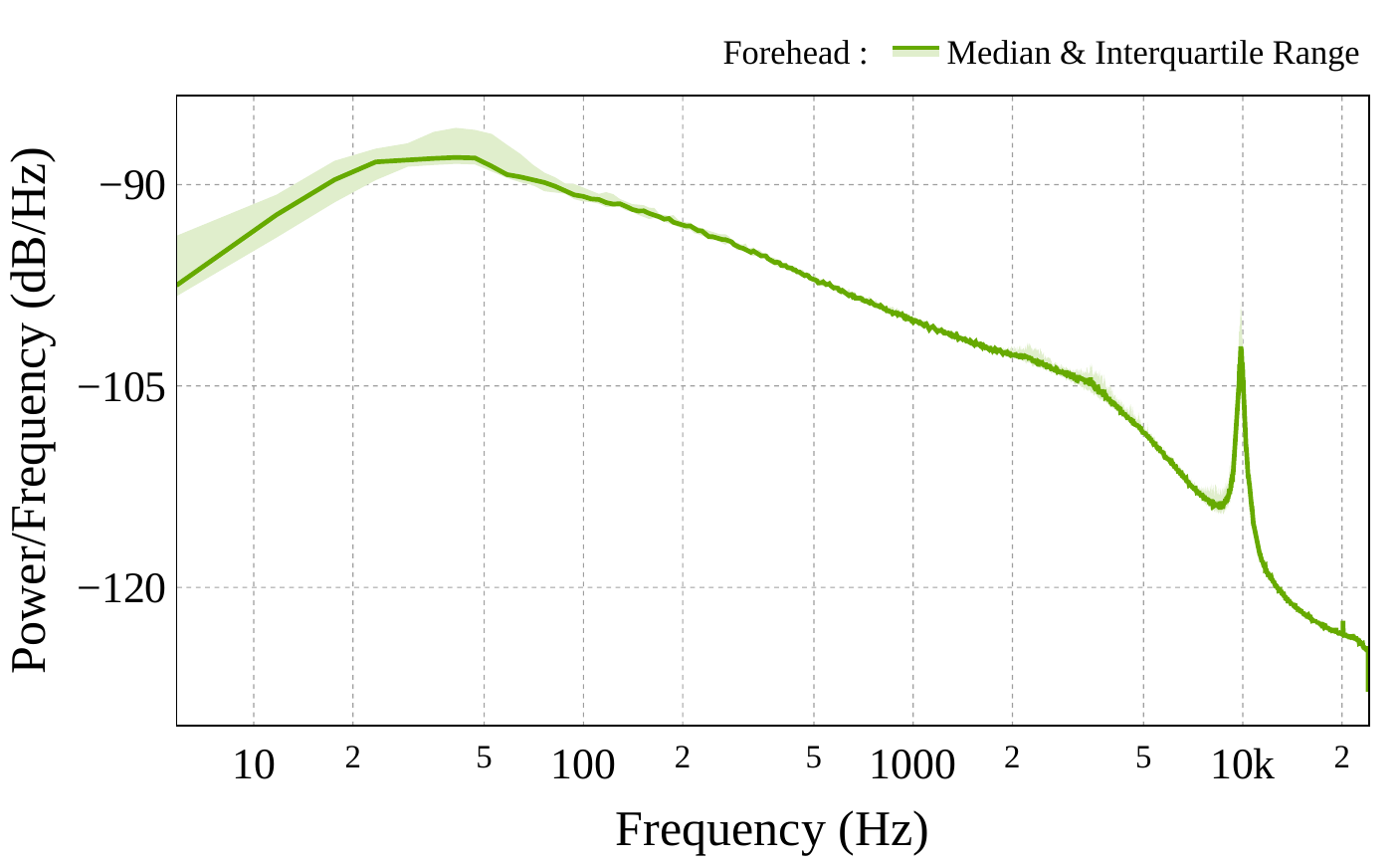}} \\
    \subfloat[Soft in-ear]{\includegraphics[width=0.49\linewidth]{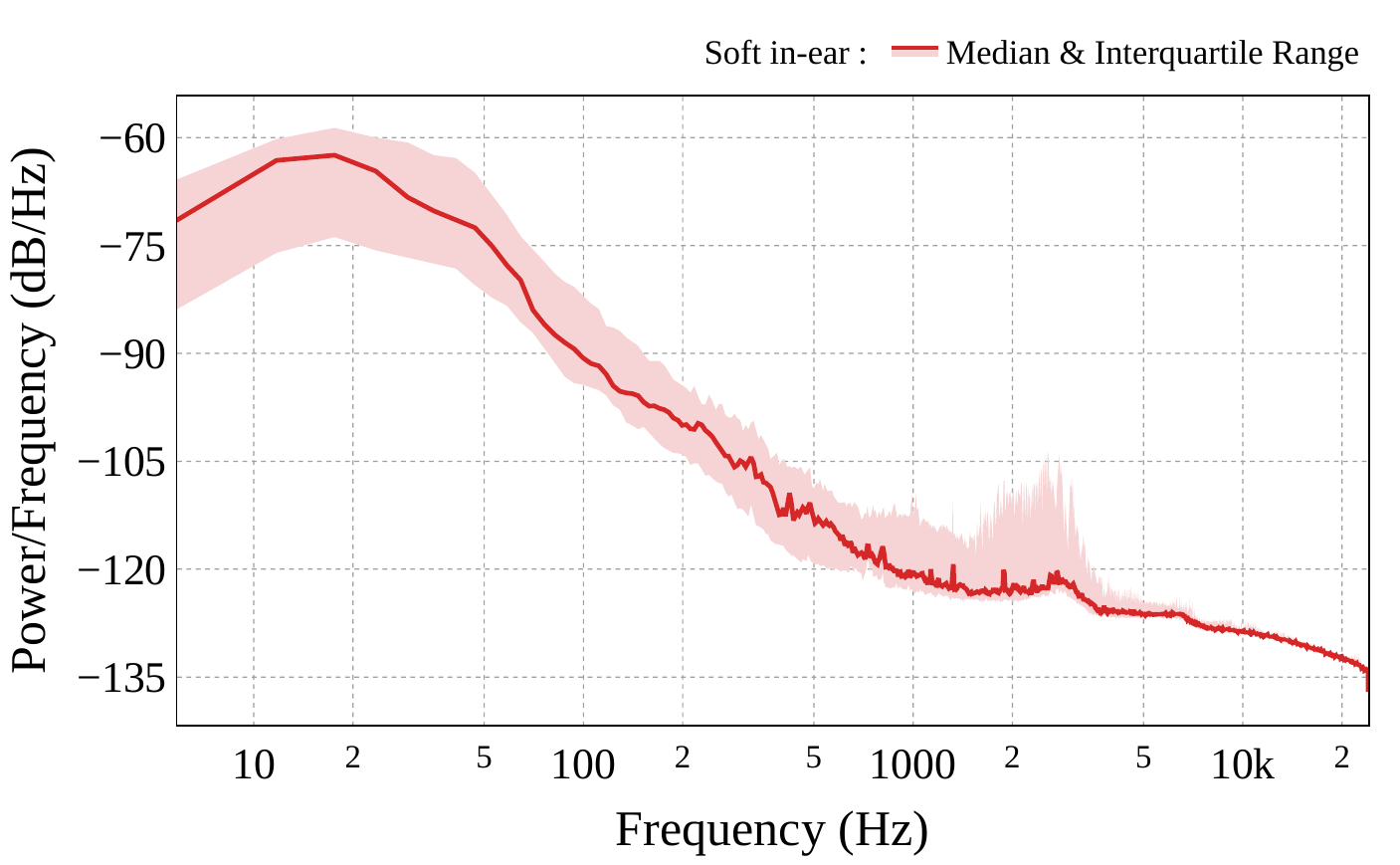}}
    \subfloat[Rigid in-ear]{\includegraphics[width=0.49\linewidth]{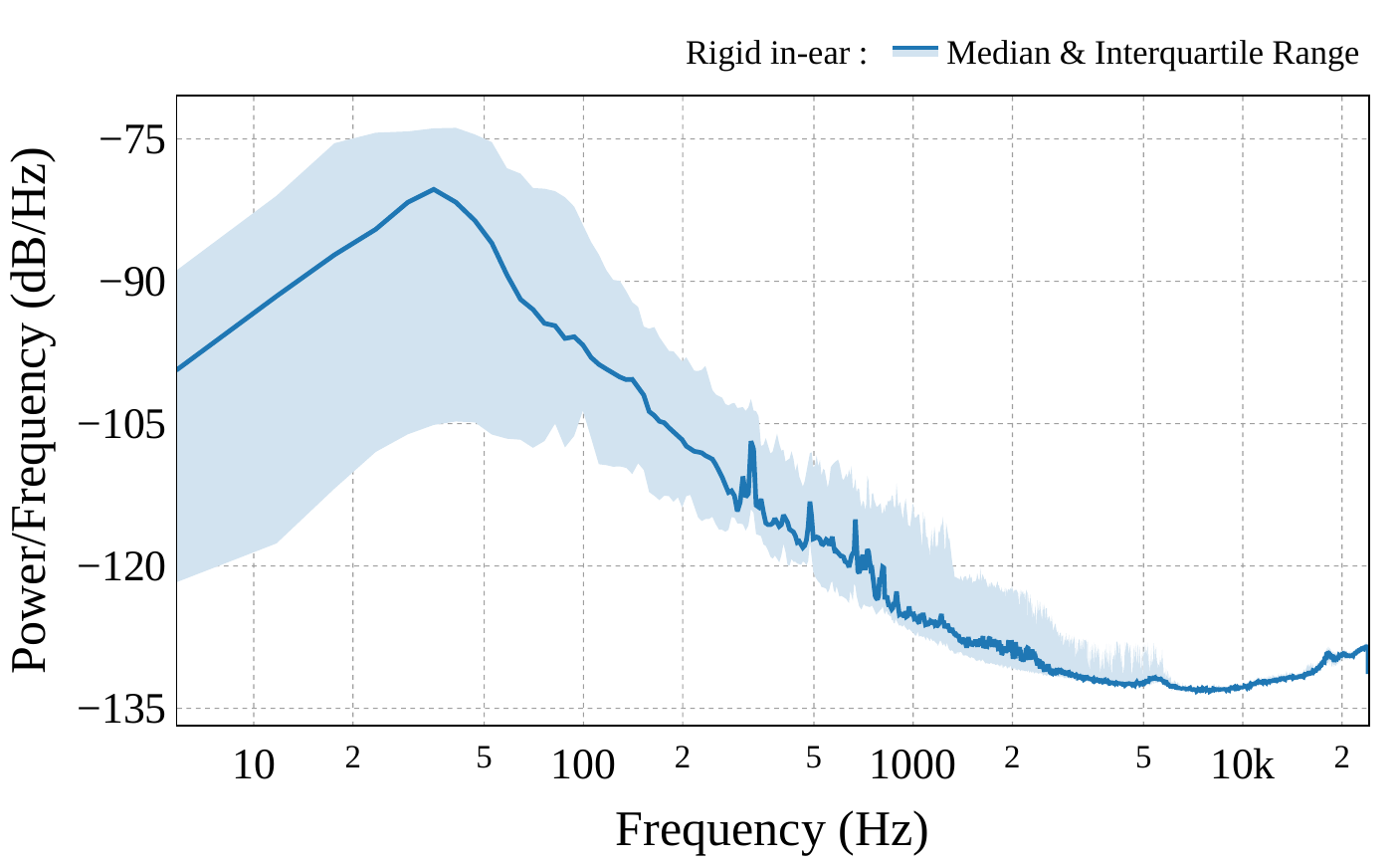}} \\
    \subfloat[Throat]{\includegraphics[width=0.49\linewidth]{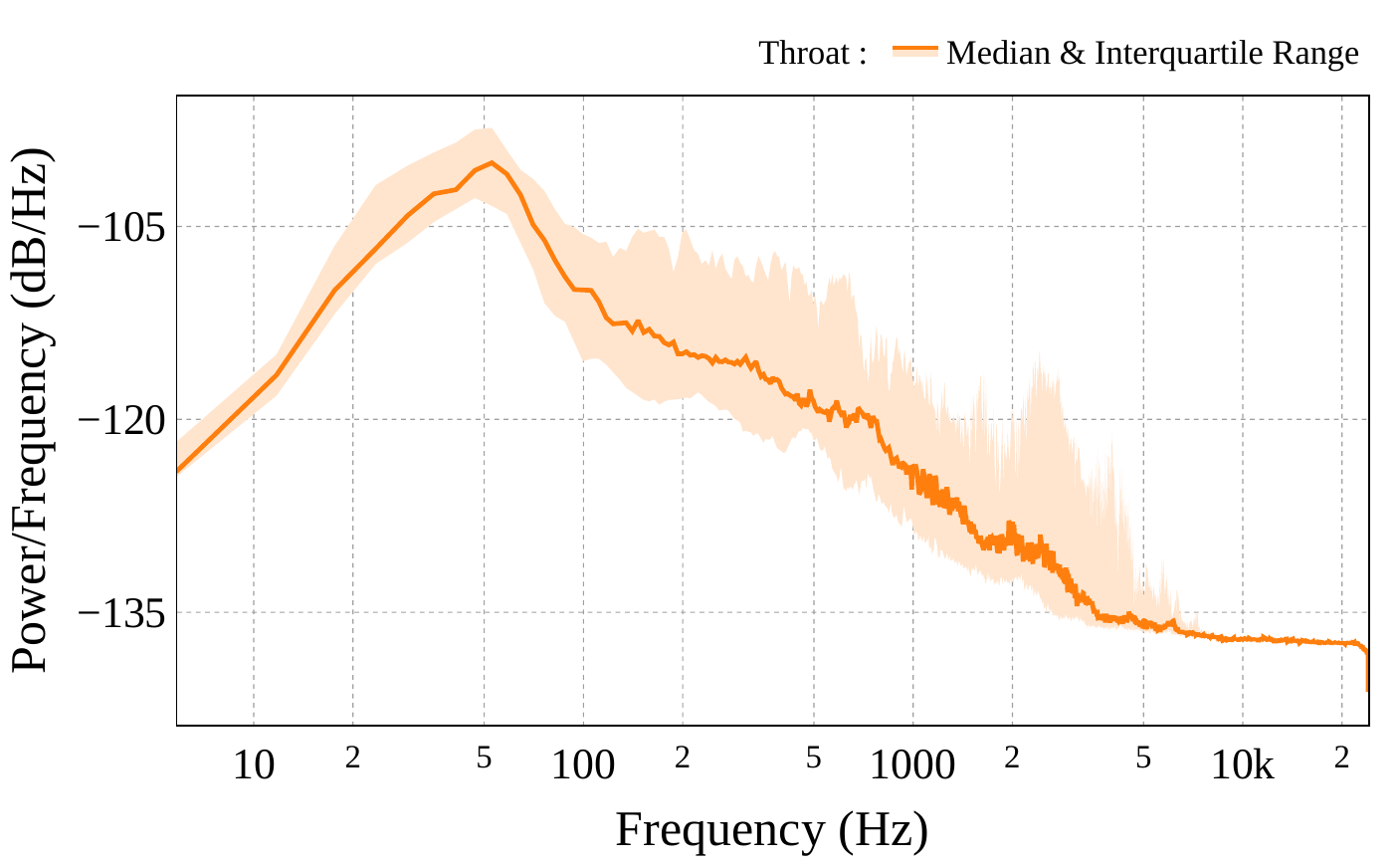}}
    \subfloat[Temple]{\includegraphics[width=0.49\linewidth]{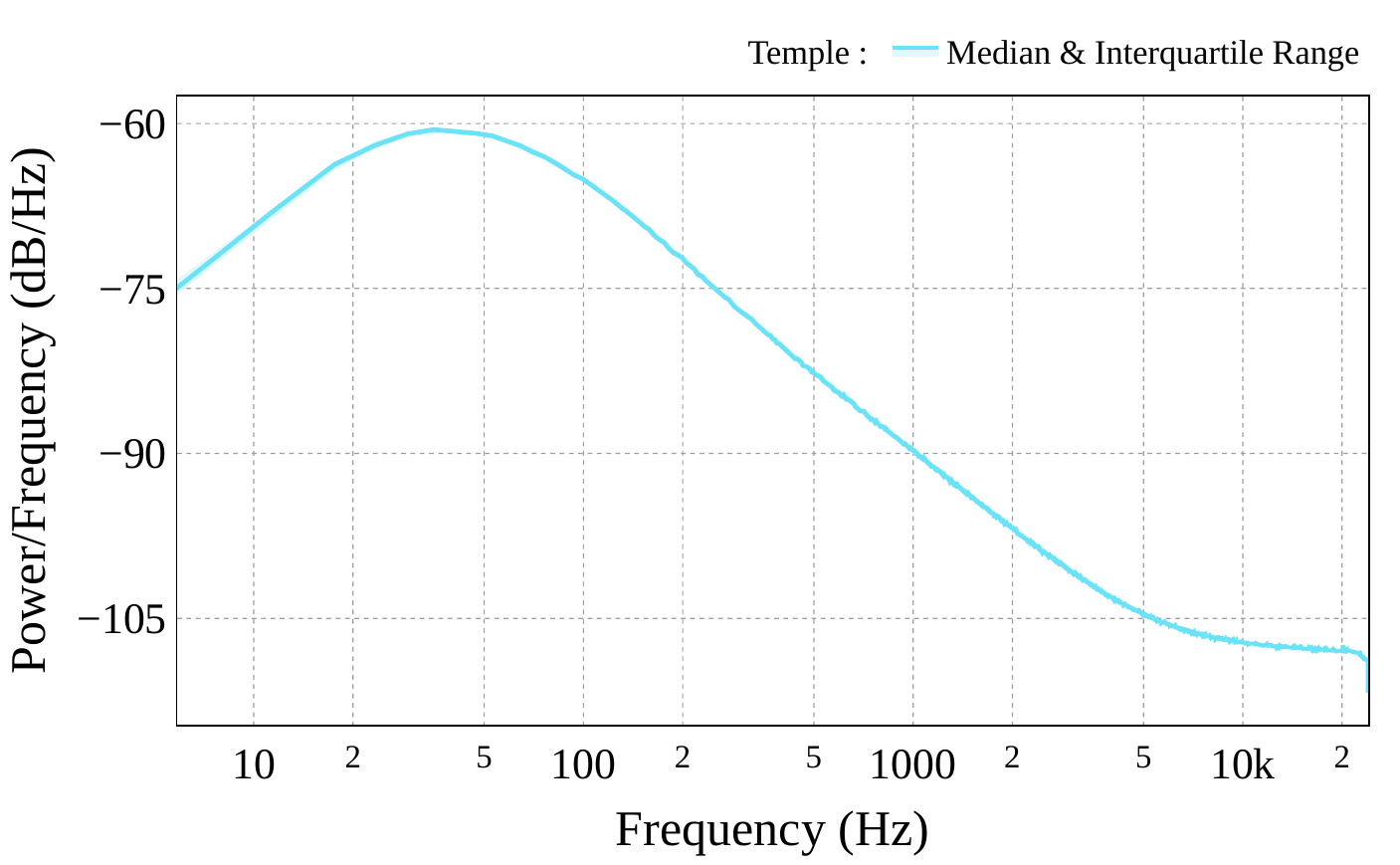}}
\caption{Median power spectral density function per sensor on \texttt{speechless-clean} given with their 25-75$\%$ IQR for all participants}
\label{fig:appendix_psd}
\end{figure*}

\newpage

\section{Speaker Verification EER on the \texttt{speech-clean} subset}

  \begin{figure}[H]
      \centering
      \subfloat[Mixed gender pairs (already given on Figure 7)]{\includegraphics[width=0.49\linewidth]{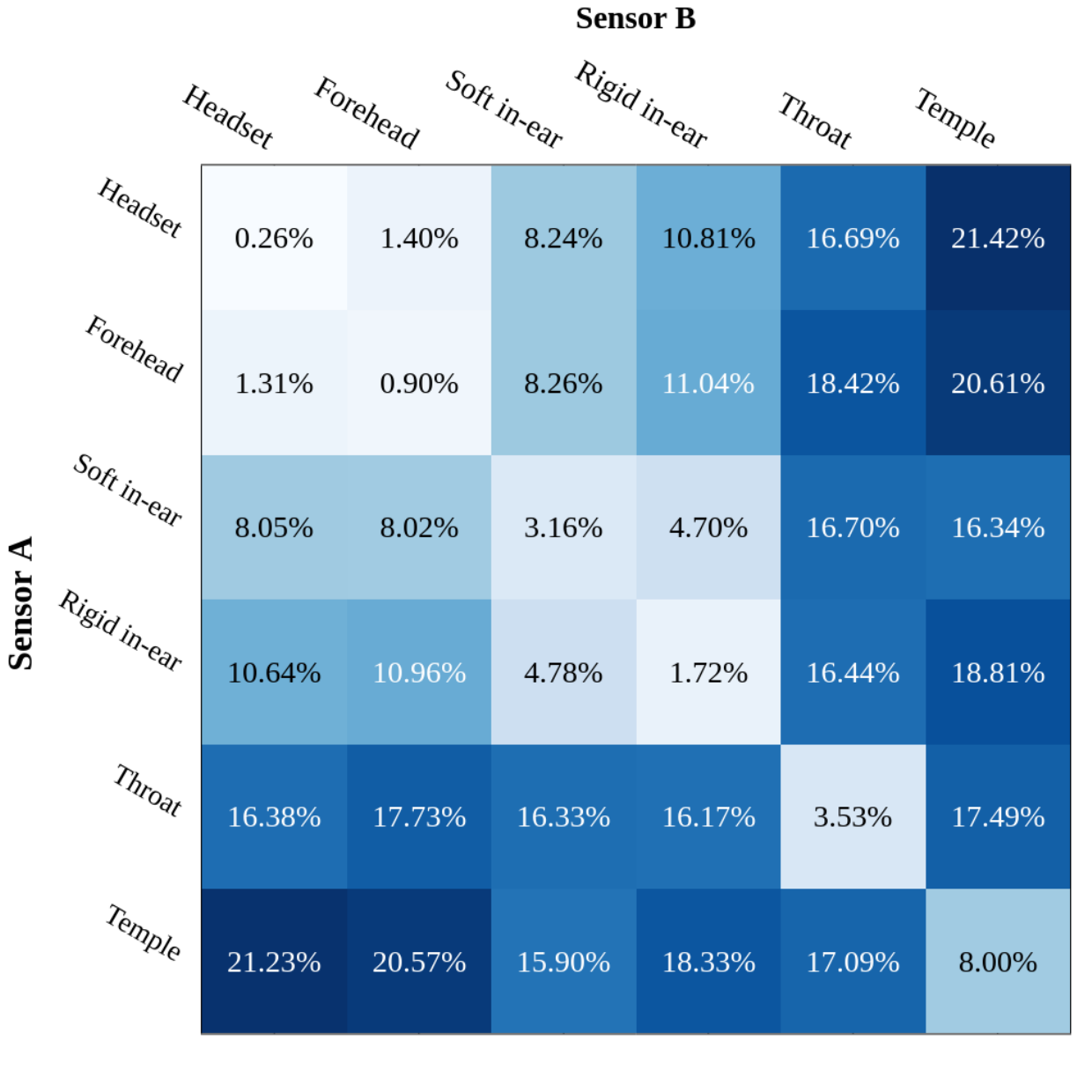}}
      \subfloat[Same gender pairs]{\includegraphics[width=0.49\linewidth]{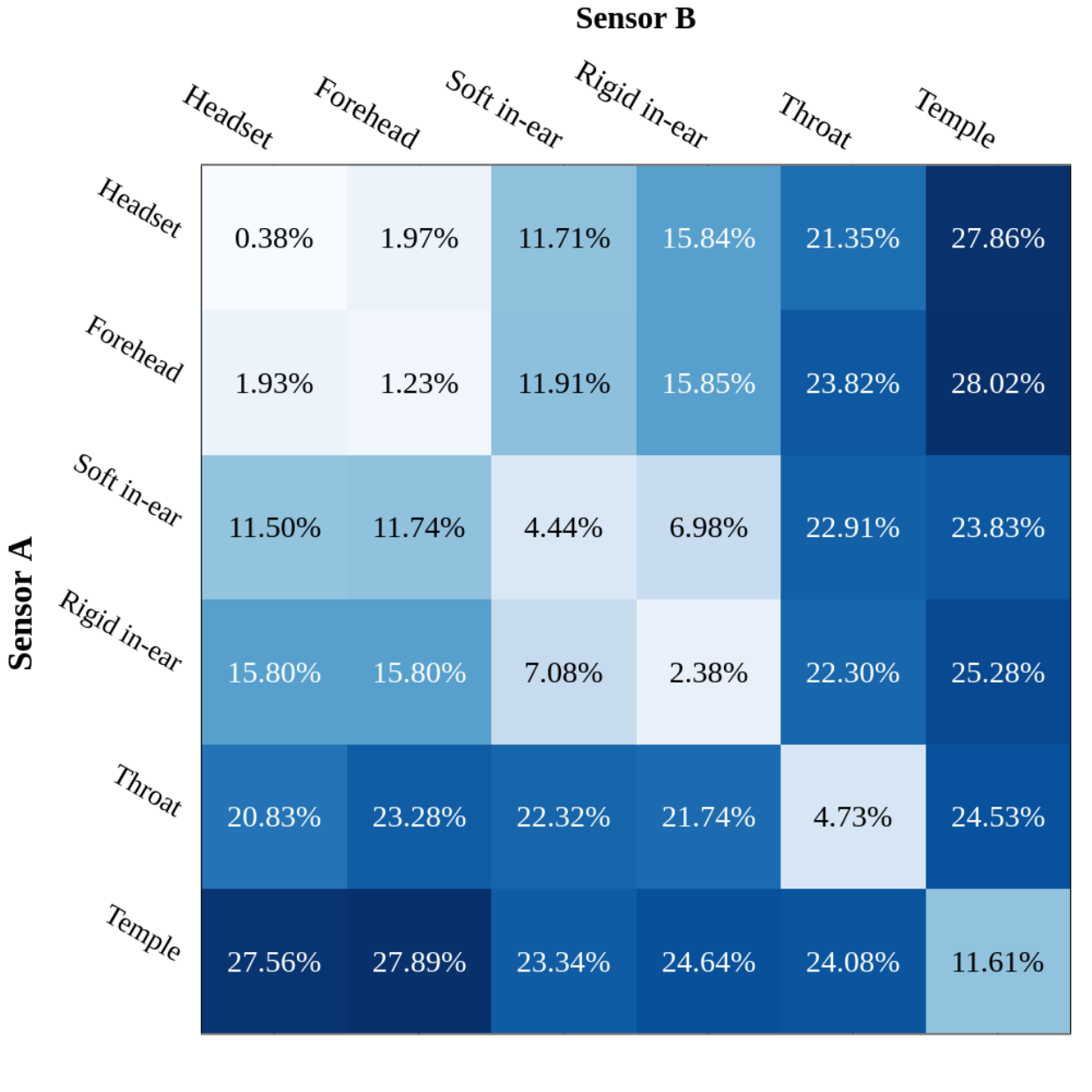}} \\
  \caption{EER obtained with the speaker verification model for sensor pairs on \texttt{speech-clean} subset}
  \label{fig:appendix_eer_raw}
\end{figure}

 \begin{figure}[H]
      \centering
      \subfloat[Mixed gender pairs]{\includegraphics[width=0.49\linewidth]{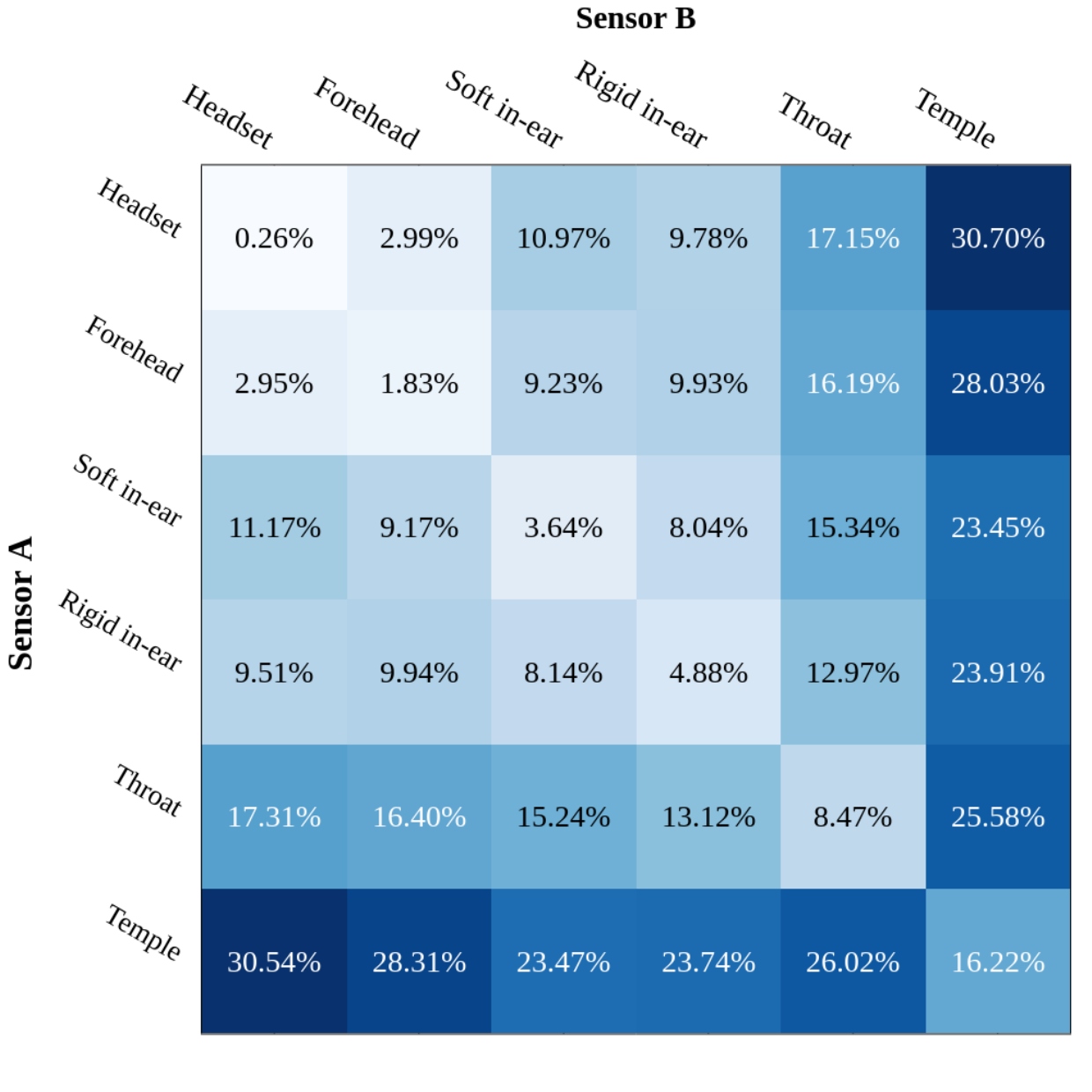}}
      \subfloat[Same gender pairs]{\includegraphics[width=0.49\linewidth]{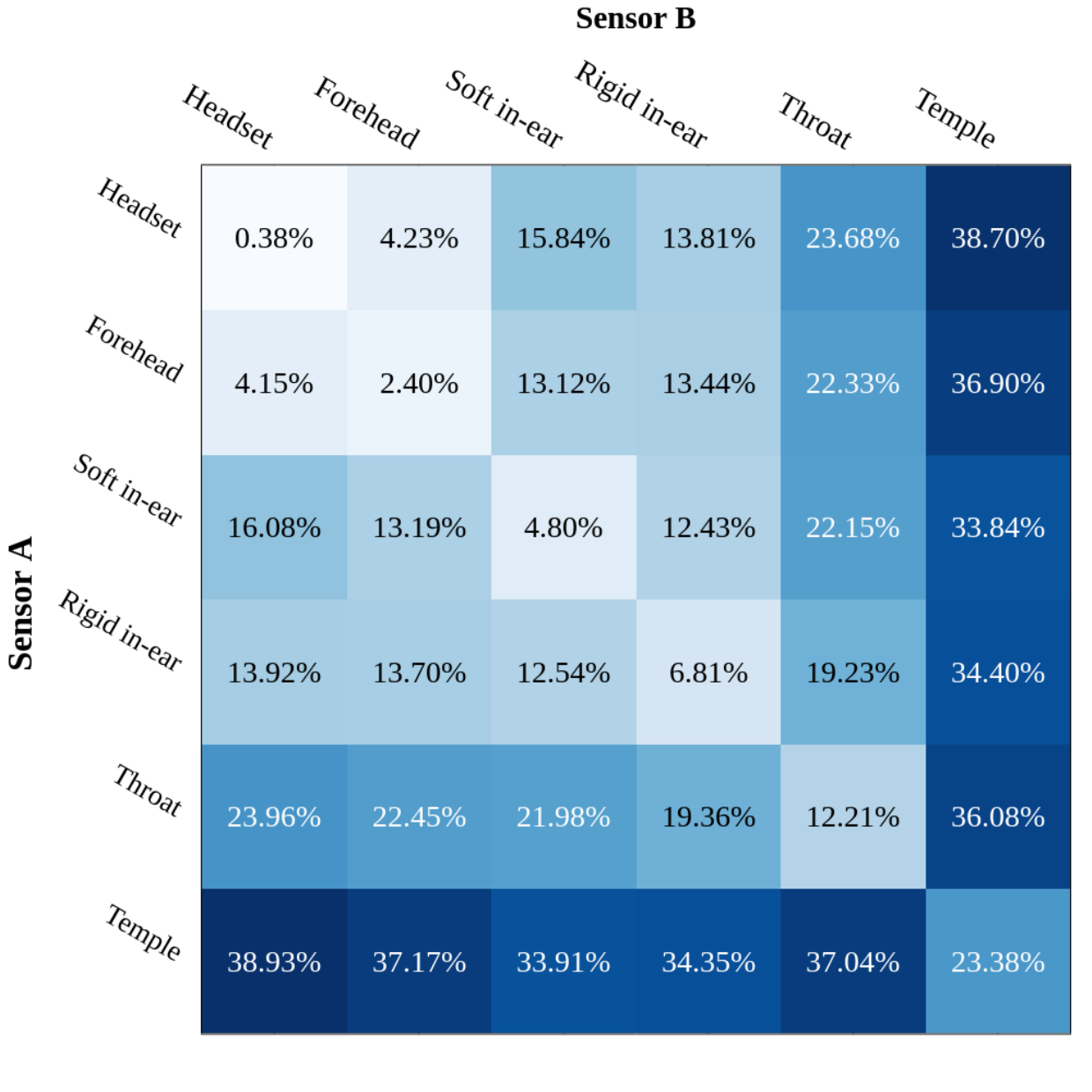}} \\
  \caption{EER obtained with the speaker verification model for sensor pairs enhanced by EBEN on \texttt{speech-clean} subset}
  \label{fig:appendix_eer_EBEN}
\end{figure}

 \section{Speaker Verification min-DCF}

 \begin{figure}[H]
      \centering
      \subfloat[Mixed gender pairs]{\includegraphics[width=0.49\linewidth]{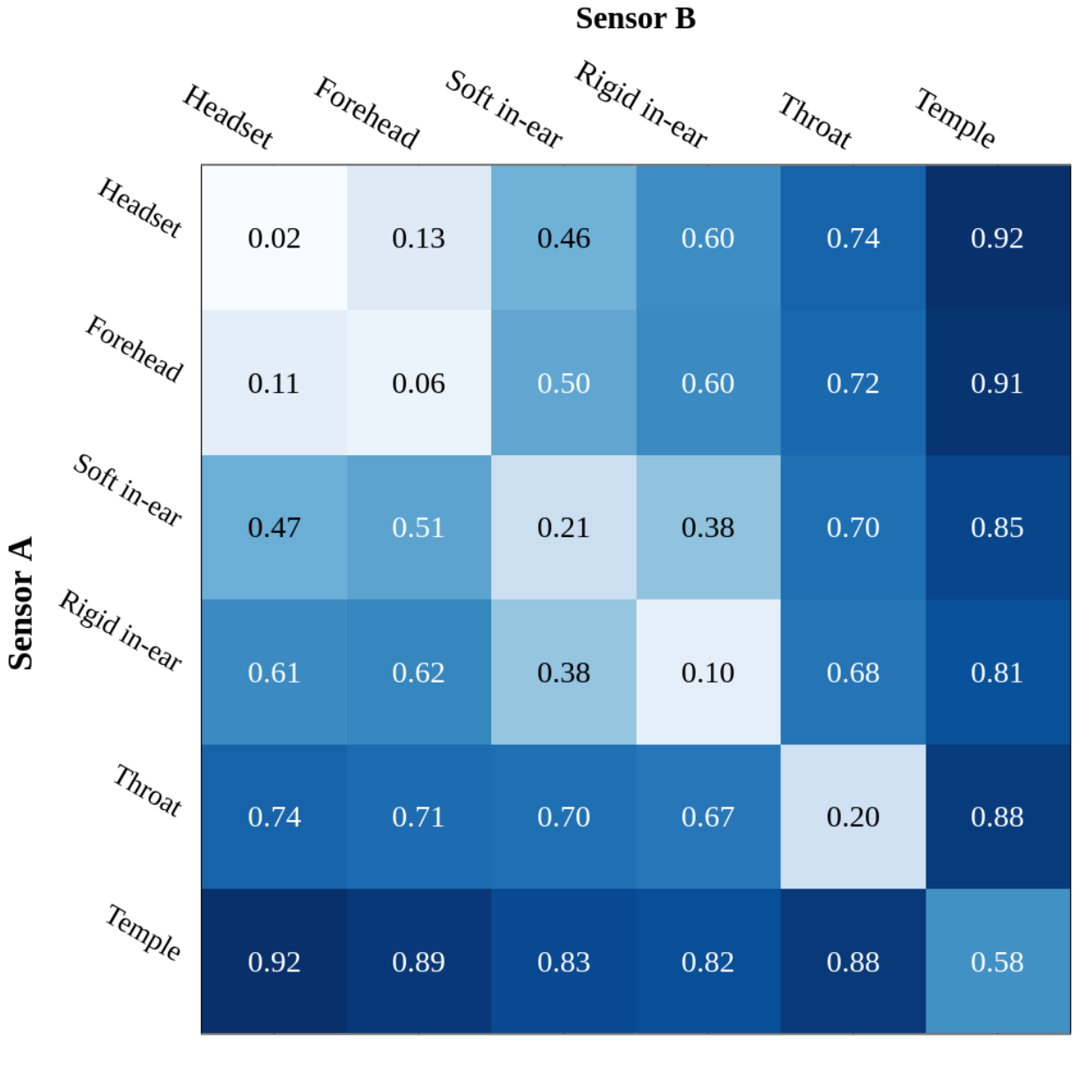}}
      \subfloat[Same gender pairs]{\includegraphics[width=0.49\linewidth]{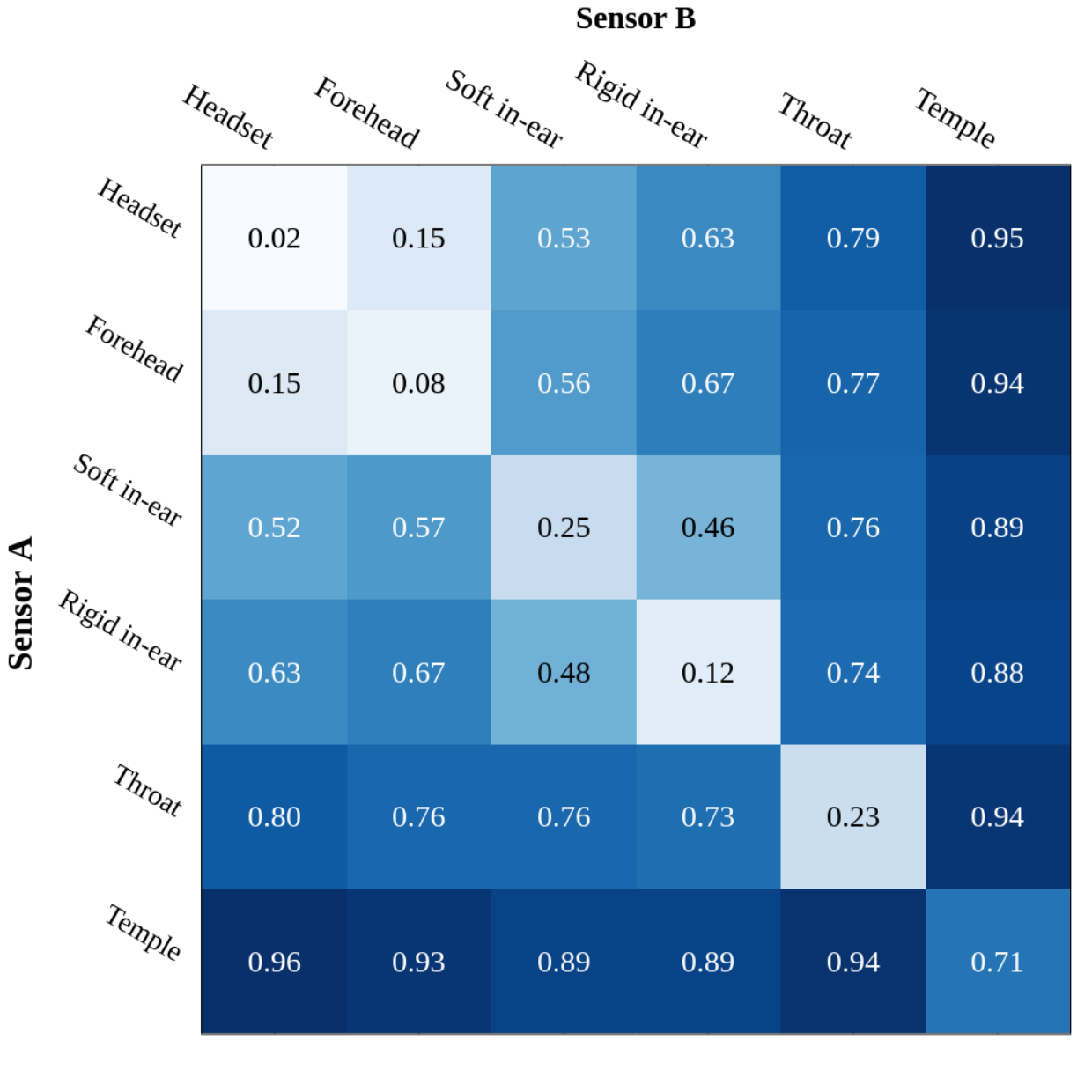}} \\
  \caption{min-DCF obtained with the speaker verification model for sensor pairs on \texttt{speech-clean} subset}
  \label{fig:appendix_dcf_raw}
\end{figure}

 \begin{figure}[H]
      \centering
      \subfloat[Mixed gender pairs]{\includegraphics[width=0.49\linewidth]{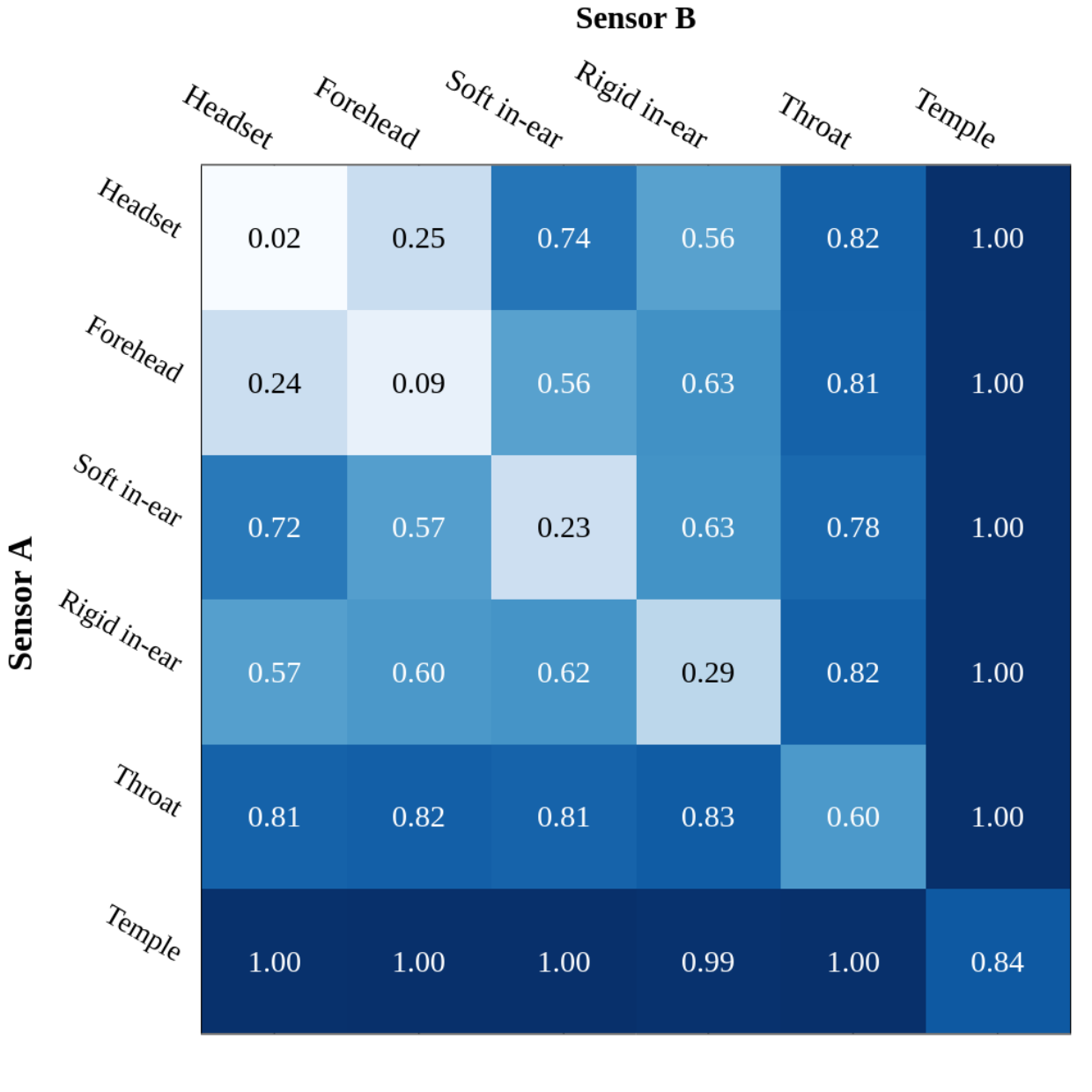}}
      \subfloat[Same gender pairs]{\includegraphics[width=0.49\linewidth]{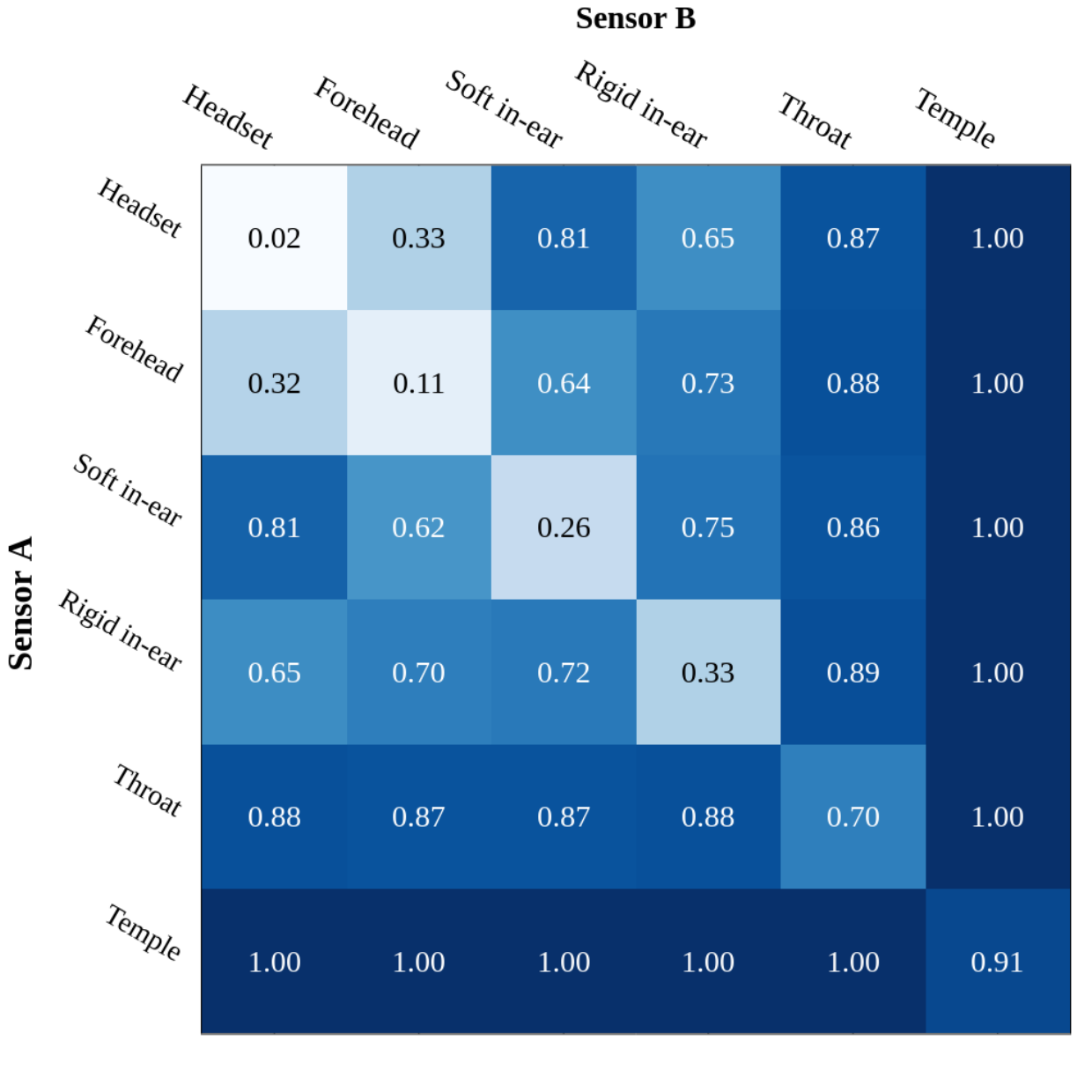}} \\
  \caption{min-DCF obtained with the speaker verification model for sensor pairs enhanced by EBEN on \texttt{speech-clean} subset}
  \label{fig:appendix_dcf_EBEN}
\end{figure}








\end{document}